\documentclass[a4paper,11pt]{article}
\usepackage[utf8]{inputenc}
\usepackage{amsmath, amssymb}
\usepackage{geometry}
\usepackage{fancyhdr}
\usepackage{authblk}  % For multiple authors and affiliations
\usepackage{hyperref}
\usepackage{cite}
\usepackage{graphicx}
\usepackage{indentfirst}
\usepackage{xcolor}
\usepackage{soul} %for strike through
\usepackage{booktabs}
\usepackage{subcaption}
\usepackage[title]{appendix}
\usepackage{siunitx}
\title{\textbf{A High-Order Immersed Boundary Method for Fluid-Structure Interaction Problems}}

\author[1,2]{Yingjie Xia}
\author[1]{Stefano Colombo}
\author[1]{David Huergo}
\author[4]{Jiaqing Kou}
\author[2]{Yuting Dai\thanks{Corresponding author: yutingdai@buaa.edu.cn}}
\author[1,3]{Esteban Ferrer}
\date{}

\affil[1]{ETSIAE-UPM-School of Aeronautics, Universidad Politécnica de Madrid, Plaza Cardenal Cisneros 3, E-28040 Madrid, Spain}
\affil[2]{School of Aeronautic Science and Engineering, Beihang University, Beijing 100191, China}
\affil[3]{Center for Computational Simulation, Universidad Politécnica de Madrid, Campus de Montegancedo, Boadilla del Monte, 28660 Madrid, Spain}
\affil[4]{School of Aeronautics, Northwestern Polytechnical University, Xi'an, China}

\begin{document}
\maketitle
\begin{abstract}
Accurate and efficient simulation of fluid-structure interaction (FSI) problems remains a central challenge in computational physics. High-order discontinuous Galerkin (DG) methods offer low numerical errors and excellent scalability on modern architectures, making them attractive for high-fidelity FSI simulations. This study presents a high-order immersed boundary method (IBM) for FSI problems which combines a volume-penalization approach with a high-order nodal DG solver. To improve near wall accuracy, an anisotropic $p$-adaptation strategy based on reinforcement learning is used to dynamically adjust the polynomial orders in the mesh elements located near the moving immersed boundaries. By doing so, we show enhanced accuracy with a limited increase in computational cost. Accurate evaluation of surface forces is achieved using symmetric high-order Gaussian quadrature on immersed boundaries. The proposed method is coupled with both rigid-body and elastic-structure solvers within a partitioned framework. Numerical validations using a pitching airfoil, stall flutter of an airfoil, and flow-induced vibration of an elastic beam behind a cylinder demonstrate high-order accuracy and robustness. These results indicate that the present approach provides an effective and scalable strategy for complex moving-boundary FSI simulations.
\end{abstract}


\tableofcontents

\section{Introduction} 

Fluid-structure interaction (FSI) is a fundamental multiphysics phenomenon that occurs in a wide range of natural, engineering, and biological systems \cite{sotiropoulos2014immersed}. Typical examples include aeroelastic deformation of aircraft wings \cite{farhat2006provably}, vibration of wind turbine blades \cite{arrigan2011control}, bridge buffeting \cite{jain1996coupled}, swimming and flying \cite{tian2014fluid}, and physiological flows in the arteries or around prosthetic valves \cite{sotiropoulos2009review}. In all of these cases, the interaction between the fluid and the structure plays a decisive role in the performance, stability, and safety of the system. For example, aeroelastic instabilities, such as flutter, can compromise structural integrity and lead to catastrophic failure \cite{xia2024stall}. As such, the ability to accurately predict FSI is critical to reducing design risks, improving reliability, and minimizing certification costs \cite{mani2023perspective, bazilevs2013computational}.

Numerical simulation has become an essential tool in FSI research, complementing theoretical analysis and experimental testing \cite{verzicco2023immersed}. It enables the exploration of complex nonlinear coupled systems that are otherwise inaccessible to analytical or experimental approaches \cite{dowell2001modeling}. With the continuous development of high-performance computing and advanced numerical algorithms, high-fidelity simulations now allow detailed and repeatable investigations of unsteady FSI phenomena \cite{hou2012numerical}, playing a central role in both fundamental research and practical engineering design.

One of the central challenges in FSI simulations lies in the accurate evaluation of the fluid forces acting on deformable structures \cite{lahooti2023dns}. This difficulty becomes especially severe in cases involving large structural displacements or separated flows, where unsteady vortices, shocks, and transition effects dominate \cite{huang2021fluid,gao2023high}. Traditional low-order CFD schemes, while robust, often suffer from excessive numerical dissipation and dispersion errors, limiting their ability to resolve fine-scale features without prohibitive mesh refinement \cite{ferrer2023high}. 
High-order numerical methods have thus emerged as a promising alternative. In particular, the high-order Discontinuous Galerkin (DG) method combines the geometric flexibility of finite elements with the locally conservative formulation of finite volumes. DG methods achieve higher accuracy and lower errors (dissipation and dispersion) for the same number of degrees of freedom as low-order methods. 
Additionally, to increase local accuracy, DG allows for $h$-adaptivity (i.e., increasing the number of elements in a particular region) or $p$-adaptivity (i.e., increasing the polynomial order in specific elements). These techniques are useful for enhancing precision in particular flow regions (e.g., near walls or in wakes); see, for example \cite{KOMPENHANS2016216, Tlales2024, Rueda-Ramírez03072023,huergo2024reinforcement}.
 % Their element-local operations also make them well-suited for parallel architectures \cite{chalmers2019parallel}. 
These characteristics make DG a powerful tool for scale-resolving FSI simulations, providing the accuracy and efficiency required to capture unsteady flow-structure interactions.

Another major challenge in FSI arises from the treatment of moving and deforming boundaries. In the context of high-order solvers, several strategies have been proposed to address moving boundaries. The FLEXI solver \cite{gao2023high} and the 3DG solver \cite{froehle2014high} adopt the Arbitrary Lagrangian-Eulerian (ALE) approach, which tracks the FSI with body-fitting moving meshes. Although the ALE approach ensures precise boundary enforcement, it suffers from mesh distortion and possible element inversion under large structural deformations. 
Within the ALE framework, high-order sliding meshes \cite{FERRER20127037, DURRWACHTER2021104825, LAHOOTI2025113734} can be used for some FSI problems (rigid body rotating motion). This technique avoids mesh distortion but requires careful treatment (e.g., over-integration) at sliding interfaces to maintain high-order accuracy, which results in costly iterations. %and generating high-quality moving meshes for complex geometries remains computationally expensive. 
Recently, a high-order Moving Reference Frame (MRF) technique has been implemented in Nektar++ \cite{moxey2020nektar++}. This approach avoids explicit remeshing and is well-suited for problems involving large structural displacements. Nevertheless, the use of non-inertial frames introduces additional mathematical complexity, especially in the presence of multiple interacting bodies or nontrivial structural kinematics.

To overcome these challenges, the immersed boundary method (IBM), which is designed inherently for moving geometries, has been adapted into high-order frameworks. IBM decouples the mesh from the geometry by imposing boundary conditions through penalization or forcing terms in the Navier-Stokes equations while solving the equations in a fixed (often Cartesian) background grid. Kou \textit{et al.} \cite{kou2022immersed} implemented IBM volume-penalization within a high-order flux reconstruction framework and explained that local adaptation near the IBM was advantageous to recover high accuracy. Based on this work, Ferrer \textit{et al.} \cite{ferrer2023high} integrated IBM into the Horses3D DG solver, extending IBM capabilities to high-order DG formulations. Subsequently, Colombo \textit{et al.} \cite{colombo2025high} introduced a surrogate boundary technique within the DG-IBM framework to alleviate instabilities associated with badly cut elements, demonstrating high-order convergence. All of these efforts were limited to fixed geometries. The extension of DG-IBM frameworks to handle moving geometries (via immersed boundaries) remains an open challenge that we address in this work.

Moving boundaries in FSI often induce sharp gradients and localized flow structures, requiring adaptive strategies to maintain accuracy while controlling computational cost. Conventional adaptive mesh refinement \cite{fraysse2012comparison} and $hp$-adaptation approaches \cite{kompenhans2016comparisons} rely on heuristic indicators or manually tuned error estimators, which can be computationally expensive and suboptimal in capturing dynamically evolving regions. Recent advances in Reinforcement Learning (RL) offer a promising avenue for automated adaptation. Huergo \textit{et al.} \cite{huergo2024reinforcement} demonstrated the potential of RL to automate anisotropic $p$-adaptation and error estimation in high-order solvers, achieving accurate solutions with reduced computational cost. However, its application to moving boundary problems remains unexplored. The present work bridges this gap by integrating RL-driven $p$-adaptation into a DG-IBM framework, enhancing local resolution near moving FSI interfaces without reducing overall efficiency.

A further challenge in FSI simulation lies in the coupling of the fluid and structural solvers. Monolithic coupling \cite{hubner2004monolithic, ryzhakov2010monolithic} solves the entire system simultaneously, providing superior stability, but at the cost of complexity and scalability. Partitioned coupling, in contrast, treats each domain separately and exchanges data at the interface in a staggered manner \cite{froehle2014high, kim2018weak}, allowing the reuse of existing solvers and flexible multi-physics integration. In this work, a partitioned coupled strategy is employed to integrate the high-order DG-IBM fluid solver with multiple structural solvers, including both rigid-body and elastic models. This modular design provides a flexible platform for investigating various FSI problems, while maintaining computational efficiency and scalability.

Despite IBM having been widely studied, its application within high-order solvers, especially for moving boundary FSI problems, remains limited. This work develops a high-order DG-IBM framework enhanced by an RL-based anisotropic $p$-adaptation strategy, enabling accurate and efficient simulations of FSI problems. To the authors’ knowledge, this is the first demonstration of a high-order DG-IBM approach with $p$-adaptation successfully applied to moving-boundary FSI problems. 

The remainder of this paper is organized as follows. Section \ref{sec:Numerical Methodology} presents the methodology, including the high-order DG-IBM framework and the $p$-adaptation strategy. Section \ref{sec:FSI} introduces the FSI coupling framework and the implementation of rigid-body and elastic structural solvers. Section \ref{sec:Test Cases} provides validation and performance analysis of the proposed approach. Finally, conclusions are drawn in Section \ref{sec:Conclusions}.

\section{Numerical Methodology}
\label{sec:Numerical Methodology}

The present work uses the open-source high-order solver HORSES3D \cite{ferrer2023high}, which is based on a nodal Discontinuous Galerkin Spectral Element Method (DGSEM). The DG formulation provides high accuracy, low numerical errors (dissipation and dispersion)\cite{MANZANERO2020104440}, and excellent scalability in modern parallel architectures, including CPUs and GPUs. This DG framework allows one to set a polynomial order $p$ to obtain $p+1$ other accuracy in the solution. Additionally, anisotropic local p-refinement can be performed in 3D meshes to improve the solution.

In what follows, we introduce the methodology necessary to perform FSI problems using DG, including the governing equations, the high-order DG discretization, the immersed boundary method with volume-penalization, the p-adaptation strategy for moving boundaries, and the evaluation of fluid forces at the immersed interface.

\subsection{Governing equations for the fluid domain}

The three-dimensional compressible non-dimensional Navier-Stokes equations can be compactly written as
\begin{equation}
\frac{\partial \mathbf{u}}{\partial t} + \nabla \cdot \overleftrightarrow{F}_e
= \nabla \cdot \overleftrightarrow{F}_v + \mathbf{s}, \label{eq:NS}
\end{equation}
where $\mathbf{u}$ is the state vector of conservative variables $
\mathbf{u} = [\rho, \rho u, \rho v, \rho w, \rho e]^{T}.$
Here, $\mathbf{s}$ denotes a volumetric source term that will later be used to impose immersed boundary conditions through volume penalization.

The inviscid fluxes are expressed as
\begin{equation}
\overleftrightarrow{F}_e =
\begin{bmatrix}
\rho u & \rho v & \rho w \\[3pt]
\rho u^2 + p & \rho uv & \rho uw \\[3pt]
\rho uv & \rho v^2 + p & \rho vw \\[3pt]
\rho uw & \rho vw & \rho w^2 + p \\[3pt]
\rho uH & \rho vH & \rho wH
\end{bmatrix},
\end{equation}
where $\rho$, $e$, $H = E + p / \rho$, and $p$ are the density, total energy, total enthalpy, and pressure, respectively. And $\mathbf{v}=[u,v,w]^T$ is the velocity. The viscous fluxes are defined as
\begin{equation}
\overleftrightarrow{F}_v =
\begin{bmatrix}
0 & 0 & 0 \\[3pt]
\tau_{xx} & \tau_{xy} & \tau_{xz} \\[3pt]
\tau_{yx} & \tau_{yy} & \tau_{yz} \\[3pt]
\tau_{zx} & \tau_{zy} & \tau_{zz} \\[3pt]
\sum_{j=1}^{3} v_j \tau_{1j} + \kappa T_x &
\sum_{j=1}^{3} v_j \tau_{2j} + \kappa T_y &
\sum_{j=1}^{3} v_j \tau_{3j} + \kappa T_z
\end{bmatrix},
\end{equation}
where $\kappa$ is the thermal conductivity, $T_x, T_y, T_z$ are the temperature gradients, and the stress tensor is given by
\begin{equation}
\tau_{ij} = \mu\left( \frac{\partial v_i}{\partial x_j} + \frac{\partial v_j}{\partial x_i} \right)
- \frac{2}{3}\mu\,\delta_{ij}\,\frac{\partial v_k}{\partial x_k},
\end{equation}
where $\mu$ is the dynamic viscosity and $\delta_{ij}$ is the Kronecker delta.

\subsection{Volume penalization immersed boundary method}

The presence of an immersed geometry is modeled by a volumetric penalization term added to the Navier-Stokes equations. An STL (Standard Tessellation Language) surface, tessellated with triangles, represents the solid geometry and is used as the mask function $\zeta(\boldsymbol{x},t)$ to distinguish the solid region $\Omega_b$ from the fluid region $\Omega_f$ (with $\Omega=\Omega_b\cup\Omega_f$):
\begin{equation}
\zeta(\boldsymbol{x},t)=
\begin{cases}
1, & \boldsymbol{x}\in\Omega_b,\\[4pt]
0, & \boldsymbol{x}\in\Omega_f.
\end{cases}
\end{equation}
It is important to note that in DG, penalized points correspond to high-order Legendre-Gauss points that define the polynomial within each mesh element, leading to a high-order representation of the immersed boundary. The mask is used to localize the penalizing source $\mathbf{s}$ in Eq.\eqref{eq:NS}, which takes the form as
\begin{equation}
\mathbf{s} \;=\; \frac{\zeta}{\eta}
\begin{bmatrix}
0\\[4pt]
\rho(u_s - u)\\[4pt]
\rho(v_s - v)\\[4pt]
\rho(w_s - w)\\[4pt]
\frac{\rho}{2}\bigl(u_s^2+v_s^2+w_s^2\bigr) - \frac{\rho}{2}\bigl(u^2+v^2+w^2\bigr)
\end{bmatrix},
\end{equation}
where $(u_s,v_s,w_s)^{\mathrm{T}}$ is the target solid velocity at the solution point and $\eta>0$ is the penalization parameter. The penalization term enforces the desired velocity inside the immersed region by strongly damping the momentum difference between fluid and solid. In practice $\eta$ must be chosen small enough to approximate the solid porosity while avoiding excessive stiffness; a common practical choice in explicit time-stepping schemes is $\eta=\Delta t$ (the time step) to balance enforcement and stability.

The volume penalization approach does not require cell cutting or body-fitted meshes and therefore is readily applied to curvilinear and unstructured element distributions; this geometric decoupling makes the method compatible with the element-local operations of the DG discretization and with high-order polynomial bases. The penalization is implemented pointwise at solution nodes, so it preserves the locality and parallel efficiency of the DGSEM. For more details on the implementation, see \cite{kou2022immersed, colombo2025high}.

In order to handle the moving boundaries, the mask is reinitialized at each time step from the updated surface geometry. The solid velocity at immersed points is obtained from the structural solver output, which will be discussed in Section \ref{sec:structural solver}. It gives a unified treatment of both rigid and flexible motion within the same DG-IBM framework.

Finally, as moving interfaces can traverse regions with coarse background mesh resolution, we combine the volume penalization with the RL-driven anisotropic $p$-adaptation to locally increase polynomial degree near the immersed boundary and also in vortical regions. The $p$-adaptation therefore complements the penalization by improving local resolution where the mask transitions occur. 

\subsection{High-order discontinuous Galerkin discretization}

The physical domain is partitioned into non-overlapping curvilinear hexahedral elements $e$, each mapped to a reference element $e_l$ by a polynomial transfinite mapping. Let $J$ denote the Jacobian of this mapping and $\nabla_{\xi}$ the differential operator in reference space. Applying the mapping to Eq.\eqref{eq:NS} induces the transformed form
\begin{equation}
J\,\mathbf{u}_t + \nabla_{\xi}\cdot\overleftrightarrow{F}_e = \nabla_{\xi}\cdot\overleftrightarrow{F}_v + J\,\mathbf{s},
\end{equation}
where $\mathbf{s}$ is the volumetric source term introduced in Section~2.1.

The weak local form is obtained by multiplying the transformed equations by a test function $\phi_i$ (taken equal to the local basis) and integrating over a reference element $e_l$:
\begin{equation}
\int_{e_l} J\,\mathbf{u}_t\,\phi_i\,\mathrm{d}V
+ \int_{\partial e_l} \phi_i\,(\overleftrightarrow{F}_e - \overleftrightarrow{F}_v)\cdot\hat{n}\,\mathrm{d}S
- \int_{e_l} \nabla_{\xi}\phi_i\cdot(\overleftrightarrow{F}_e - \overleftrightarrow{F}_v)\,\mathrm{d}V
= \int_{e_l} J\,\mathbf{s}\,\phi_i\,\mathrm{d}V,
\end{equation}
where $\hat{n}$ is the outward unit normal on $\partial e_l$. To couple neighboring elements, discontinuous fluxes on element faces are replaced by numerical fluxes. The inviscid numerical flux $\overleftrightarrow{F}_e^{\ast}$ is computed with an upwind Riemann solver, and in this work we use the Lax-Friedrichs flux. Viscous numerical fluxes $\overleftrightarrow{F}_v^{\ast}$ are expressed in terms of both state and gradient traces, and in this work, viscous discretization is performed using the Bassi-Rebay 1 scheme.

Summing the local contributions over all elements and replacing face fluxes by $\overleftrightarrow{F}_e^{\ast}$ and $\overleftrightarrow{F}_v^{\ast}$ induces the global semi-discrete weak form
\begin{equation}
\sum_{e_l}\int_{e_l}\bigl[J\,\mathbf{u}_t\,\phi_i - \nabla_{\xi}\phi_i\cdot(\overleftrightarrow{F}_e - \overleftrightarrow{F}_v) - J\,\mathbf{s}\,\phi_i\bigr]\,\mathrm{d}V
+ \sum_{\partial e_l}\int_{\partial e_l}\phi_i\,(\overleftrightarrow{F}_e^{\ast}-\overleftrightarrow{F}_v^{\ast})\cdot\hat{n}\,\mathrm{d}S = 0.
\end{equation}

The final step, constructing a usable numerical scheme, is to approximate the solution and fluxes with polynomials of order 
$p$, and to evaluate the volume and surface integrals using Gaussian quadrature. In HORSES3D, both Gauss-Legendre and Gauss-Lobatto quadrature points are supported, although only Gauss-Legendre points are used in this work. Non-conforming elements are coupled through the mortar method \cite{ferrer2023high,10.1063/5.0241311}. %Further details of the implementation can be found in \cite{ferrer2023high}.

\subsection{p-adaptation for moving immersed boundary}

Local $p$-refinement provides an efficient mechanism to improve accuracy near complex boundary layers without modifying the underlying mesh. Due to the flexibility of the high-order DG framework, the degree of polynomial $p$ within each element can be varied locally, leading to substantial computational savings compared to uniform high-order discretizations. Furthermore, the polynomial adaptation is anisotropic, allowing us to set three different polynomial orders inside each element, one for each axis $x$, $y$, and $z$. This feature is particularly beneficial in the vicinity of moving immersed boundaries.

In this work, we employ an automated anisotropic $p$-adaptation strategy based on an RL framework\cite{huergo2024reinforcement}. In this approach, the adaptation process is formulated as a Markov decision process, in which an RL agent dynamically adjusts the local polynomial degree according to the estimated solution accuracy, local flow features, and moving boundary. The agent state represents the flow field, defined as a set of variables evaluated at every Gauss node inside an element of the computational mesh (in the reference space). 
%These variables can be selected by the user and must be representative of the problem to be solved. 
We typically use the normalized momentum $(\rho u, \rho v, \rho w)$, which has proven to provide good results for a variety of problems. The action space consists of three discrete operations for each spatial direction: increase $(+1)$, maintain $(0)$, or decrease $(-1)$ the current polynomial degree. The reward function balances accuracy and computational efficiency through a tradeoff between the predicted local error and the associated cost:
\begin{equation}
r = \underbrace{\left(\frac{p_{\max}}{p}\right)^{\alpha}}_{\text{COST}}
\underbrace{\exp\!\left[-\frac{\mathrm{RMSE}^2}{2\sigma^2}\right]}_{\text{ERROR}},
\end{equation}
where $\alpha=0.9$ and $\sigma=0.05$ are empirically tuned parameters. The optimal policy is obtained through a value-iteration algorithm and deployed in the solver as an offline-trained agent. During simulations, the adaptation is periodically triggered (every $10^2\!\sim\!10^3$ time steps), allowing each element to adjust its polynomial degree $p\in[1,6]$ for each coordinate axis, with a maximum change of one order per update to ensure numerical stability.

The novelty of the present work lies in extending this RL-based anisotropic $p$-adaptation strategy to problems that involve moving immersed boundaries. When the immersed geometry moves or deforms, the spatial distribution of the mask function $\zeta(\mathbf{x},t)$ and the associated flow gradients evolve rapidly, leading to significant variations in local resolution requirements. To maintain accuracy and efficiency, the RL agent dynamically reallocates the polynomial order near the time-dependent interface by responding to changes in the local flow field and penalization source term $\mathbf{s}$. This enables accurate capture of unsteady boundary-layer development and wake dynamics without remeshing or ad-hoc refinement criteria.

The combined DG-IBM-RL framework therefore achieves a self-adaptive resolution capability: the immersed boundary is represented implicitly through the penalization term, while the surrounding numerical resolution is automatically tuned by the RL-driven anisotropic $p$-adaptation. This hybridization ensures both robustness for moving boundaries and computational efficiency in large-scale unsteady flow simulations.

\subsection{Fluid force evaluation on immersed boundary}
\label{sec:fluid force}
An accurate evaluation of the forces of the fluid at the immersed boundary is essential for reliable FSI simulations. In the present work, the solid surface, i.e., immersed boundary, is discretized into triangular elements derived from the STL geometry. The surface normals and element areas are computed directly from this discretization. The velocity and pressure fields at the element integration points are reconstructed using an inverse distance weighting (IDW) method from the surrounding fluid solution points. Specifically, the interpolated value at each integration point on the solid surface element is given by
\begin{equation}
\mathbf{u}_{\rm IP} = \frac{\sum_{i=1}^{N_{\rm IP}} \mathbf{u}_i / d_i}{\sum_{i=1}^{N_{\rm IP}} 1/d_i},
\end{equation}
where $\mathbf{u}_i$ is the conservative variable at the $i$-th neighboring fluid point and $d_i$ is its distance to the interpolation point. $N_{\rm IP}$ is the number of interpolation points, and the effect of this value is tested and will be discussed in Appendix \ref{sec:Appendix-A}, where it is shown that the number of interpolation points has little effect on the forces for $N_{\rm IP}\in[5,25]$. The force contributions in each STL triangle are computed as
\begin{equation}
\mathbf{F}_{\rm tri} = \sum_{k=1}^{N_{\rm GP}} w_k \big( -p_k \mathbf{n}_k + \boldsymbol{\tau}_k \cdot \mathbf{n}_k \big) A_k,
\end{equation}
where $p_k$ and $\boldsymbol{\tau}_k$ are the interpolated pressure and stress tensor at the $k$-th quadrature point, $\mathbf{n}_k$ is the surface normal, $w_k$ is the quadrature weight, $A_k$ is the associated elemental area, and $N_{\rm GP}$ is the number of quadrature points. 

This combination of symmetric Gaussian quadrature and IDW-based reconstruction ensures that the calculated fluid forces are accurate and consistent with high-order spatial discretization, which provides the foundation for the implementation of high-order FSI.

\section{Fluid-Structure Interaction Framework}
\label{sec:FSI}
\subsection{Numerical method for the motion of rigid and elastic bodies}
\label{sec:structural solver}

\subsubsection*{(a) Rigid-body dynamics}

A rigid-body solver is first considered for this FSI framework, and the test results are shown in Section \ref{sec: test case - Stall flutter}. The FSI behavior of flow over a torsional elastically supported airfoil (also known as stall flutter \cite{xia2024stall, menon2019flow}) is studied. The structural system of the airfoil is governed by the following equation
\begin{equation}
I_{\mathrm{ea}}\ddot{\alpha} + D_{\mathrm{ea}}\dot{\alpha} + K_{\mathrm{ea}}(\alpha - \alpha_e) = M_{\mathrm{ea}}(t),
\end{equation}
where $\alpha$ is the angle of attack, $I_{\mathrm{ea}}$ is the moment of inertia about the elastic axis, $D_{\mathrm{ea}}$ is the damping coefficient, $K_{\mathrm{ea}}$ is the torsional stiffness, and $M_{\mathrm{ea}}(t)$ is the fluid-induced moment evaluated from the immersed boundary. The equilibrium pitch angle is denoted by $\alpha_e$.

The structural governing equation is solved analytically
at each time step. Taking the angle of attack $\alpha^n$ and the pitch velocity $\dot{\alpha}^n$ as
an initial condition, the analytical solution of $\alpha^{n+1}$ and $\dot{\alpha^{n+1}}$ at the time step $n+1$ can be calculated as follows:
\begin{equation}
\begin{aligned}
\alpha^{n+1} &= e^{\lambda\Delta t}\left[ c_1\cos(\mu\Delta t) + c_2\sin(\mu\Delta t)\right] + \frac{M_{\mathrm{ea}}^{n+1}}{K_{\mathrm{ea}}} + \alpha_e, \\
\dot{\alpha}^{n+1} &= e^{\lambda\Delta t}\left[ (c_1\lambda + c_2\mu)\cos(\mu\Delta t) + (-c_1\mu + c_2\lambda)\sin(\mu\Delta t) \right],
\end{aligned}
\end{equation}
with
\begin{equation}
\lambda=-\frac{D_{\mathrm{ea}}}{2I_{\mathrm{ea}}}
\end{equation}
\begin{equation}
\mu = \frac{\sqrt{4K_{\mathrm{ea}}{I_{\mathrm{ea}}}-D_{\mathrm{ea}}^2}}{2I_{\mathrm{ea}}}
\end{equation}
\begin{equation}
c_1 = \alpha^n - \frac{M_{\mathrm{ea}}^{n+1}}{K_{\mathrm{ea}}} - \alpha_e, 
\end{equation}
\begin{equation}
c_2 = \frac{\dot{\alpha}^n - \lambda c_1}{\mu},
\end{equation}
where $\Delta t$ is the size of the timestep.
The IBM-required displacement of the immersed boundary elements and the velocity of the immersed solid points are directly obtained from $\alpha_{n+1}$ and $\dot{\alpha}_{n+1}$.

\subsubsection*{(b) Elastic-body dynamics}

An elastic-body solver is then considered in this FSI framework, and the results are shown in Section \ref{sec: test case - elastic beam}. To account for elastic bodies dynamics, a numerical approximation by means of a Finite Element Model is made, discretizing the structure in a finite number of physical coordinates \cite{hjelmstad2022fundamentals, van2001structural}. The equation of motion describing this approximated system is given by
% we use a general structural dynamic model \cite{paz2019structural, hjelmstad2022fundamentals}. The results are shown in Section \ref{sec: test case - elastic beam}. Elastic body dynamics are governed by
\begin{equation}
\mathbf{M}\ddot{\mathbf{w}} + \mathbf{C}\dot{\mathbf{w}} + \mathbf{K}\mathbf{w} = \mathbf{F}(t),\label{eq:elasticbody governing eq}
\end{equation}
where $\mathbf{w}$ denotes the nodal displacement vector, and $\mathbf{M}$, $\mathbf{C}$, and $\mathbf{K}$ are the body mass, damping, and stiffness matrices, respectively. The external force vector $\mathbf{F}(t)$ represents the fluid-induced loads evaluated from the immersed boundary.

To generalize the formulation, we adopt a modal decomposition in which the displacement field is expanded using a modal basis
\begin{equation}
\mathbf{w}(t) = \sum_{i=1}^{N_m} q_i(t)\boldsymbol{\Phi}_i = \boldsymbol{\Phi}\mathbf{q}(t),
\end{equation}
where $\boldsymbol{\Phi}_i$ is the $i$-th mode shape and $q_i(t)$ the corresponding generalized coordinate and $N_m$ denotes the number of modes. Substitution into Eq.\eqref{eq:elasticbody governing eq} and premultiply with $\boldsymbol{\Phi}^T$ results in the decoupled modal equations

\begin{equation}
\ddot{\mathbf{q}} + 2\boldsymbol{\Xi}\boldsymbol{\Omega}\dot{\mathbf{q}} + \boldsymbol{\Omega}^2\mathbf{q} = \mathbf{f}(t),\label{eq:modal-ode}
\end{equation}
where $\boldsymbol{\Omega} = \mathrm{diag}[\omega_1, \omega_2, \dots]$, $\boldsymbol{\Xi} = \mathrm{diag}[\xi_1, \xi_2, \dots]$, $\mathbf{f}(t)=\boldsymbol{\Phi}^T\mathbf{F}(t)$, and the modal properties follow from the modal orthogonality relations
\begin{equation}
\boldsymbol{\Phi}^T \mathbf{M} \boldsymbol{\Phi} = \mathbf{I}, \qquad
\boldsymbol{\Phi}^T \mathbf{C} \boldsymbol{\Phi} = 2\boldsymbol{\Xi}\boldsymbol{\Omega, \qquad
\boldsymbol{\Phi}^T \mathbf{K} \boldsymbol{\Phi} = \boldsymbol{\Omega}^2}.
\end{equation}

For time integration, the second-order system \eqref{eq:modal-ode} is rewritten as a first-order state-space system by defining the structural state vector $\mathbf{Q}=[\mathbf{q},\dot{\mathbf{q}}]^T$. The structural equations in state-space form can be expressed as
\begin{equation}
\dot{\mathbf{Q}} =
\begin{bmatrix}
\mathbf{0} & \mathbf{I} \\
-\boldsymbol{\Omega}^2 & -2\boldsymbol{\Xi}\boldsymbol{\Omega}
\end{bmatrix}\mathbf{Q} +
\begin{bmatrix}
\mathbf{0} \\ \mathbf{I}
\end{bmatrix}\mathbf{f}(t).
\end{equation}
A fourth-order Runge-Kutta method is used to integrate the system.
The physical displacement and velocity can be easily reconstructed as
\begin{equation}
\mathbf{w} = \boldsymbol{\Phi}\mathbf{q}, \qquad
\dot{\mathbf{w}} = \boldsymbol{\Phi}\dot{\mathbf{q}},
\end{equation}
which can be easily incorporated into the solver to obtain the displacement of the tessellated immersed boundary and the velocity of the immersed solid points required by the IBM.

\subsection{Coupling strategy}

The interaction between the fluid and the structure is achieved through a partitioned coupling strategy that allows flexible integration of different structural solvers within a unified computational framework. Information exchange between the fluid and structural subsystems occurs through two coupling interfaces.

The fluid-to-structure load transfer computes the fluid forces acting on the immersed body surface, including both pressure and viscous contributions. These loads are evaluated at the immersed boundary using the Gaussian quadrature method described in Section \ref{sec:fluid force}, and subsequently transferred to the structural solvers as external fluid loads.
The structure-to-fluid information transfer provides the motion data from the structural solver to the flow solver. In the present DG-IBM framework, the immersed boundary displacement is used to update the surface representation (STL triangular elements), while the velocity of all immersed solid points (mask = 1) is employed to impose the penalization source term in the Navier--Stokes equations. These quantities are obtained from the rigid-body and elastic solvers described in Section \ref{sec:structural solver}.

A partitioned weak-coupling scheme is adopted for temporal integration. At each time step, the structural motion is updated first on the basis of the fluid loads from the previous fluid solution. The updated displacements and velocities are then passed to the fluid solver to advance the flow field to the current time level. This explicit coupling sequence continues until the end of the simulation. Although the present implementation employs weak coupling, the framework can be readily extended to a strong coupling formulation if enhanced coupling accuracy or stabilization is required.

% \subsection{Implementation overview and algorithmic workflow}

\section{Test Cases}
\label{sec:Test Cases}

The proposed high-order DG-IBM framework for FSI problems is validated through a series of benchmark problems covering prescribed, rigid-body, and elastic motions. The first case, flow past a pitching airfoil, is designed to assess the accuracy of immersed-boundary force evaluation and the efficiency of the $p$-adaptation strategy under moving boundaries. The second case, stall flutter of an airfoil, examines the predictive capability of the partitioned coupling strategy for rigid-body FSI. The third case, flow-induced vibration of an elastic beam behind a cylinder, shows the robustness and accuracy of the solver in handling flexible structure deformations.

\subsection{Flow past a pitching airfoil}
\label{sec:Case1P}

To examine the accuracy and robustness of the proposed DG-IBM framework for moving-boundary problems, a NACA0012 airfoil undergoing large-amplitude sinusoidal pitching motion is considered. This case simultaneously evaluates the immersed boundary force integration, the moving-boundary treatment under strong geometric deformation, and the performance of the RL-driven $p$-adaptation strategy for moving-immersed boundary. The prescribed motion is
\begin{equation}
\alpha(t)=\alpha_0 + A_\alpha\sin(2\pi f_\alpha time^*),
\end{equation}
with $\alpha_0=15.8^\circ$, $A_\alpha=24.4^\circ$, and nondimensional frequency $f_\alpha=0.156$, $time^*$ is nondimensional time. The airfoil rotates about its mid-chord axis. The flow conditions are $Re=1000$, $Ma=0.25$ and chord $c=1$. The CFL number is 0.3.

\begin{figure}[htbp] % htbp
  \centering
  \includegraphics[width=0.7\textwidth]{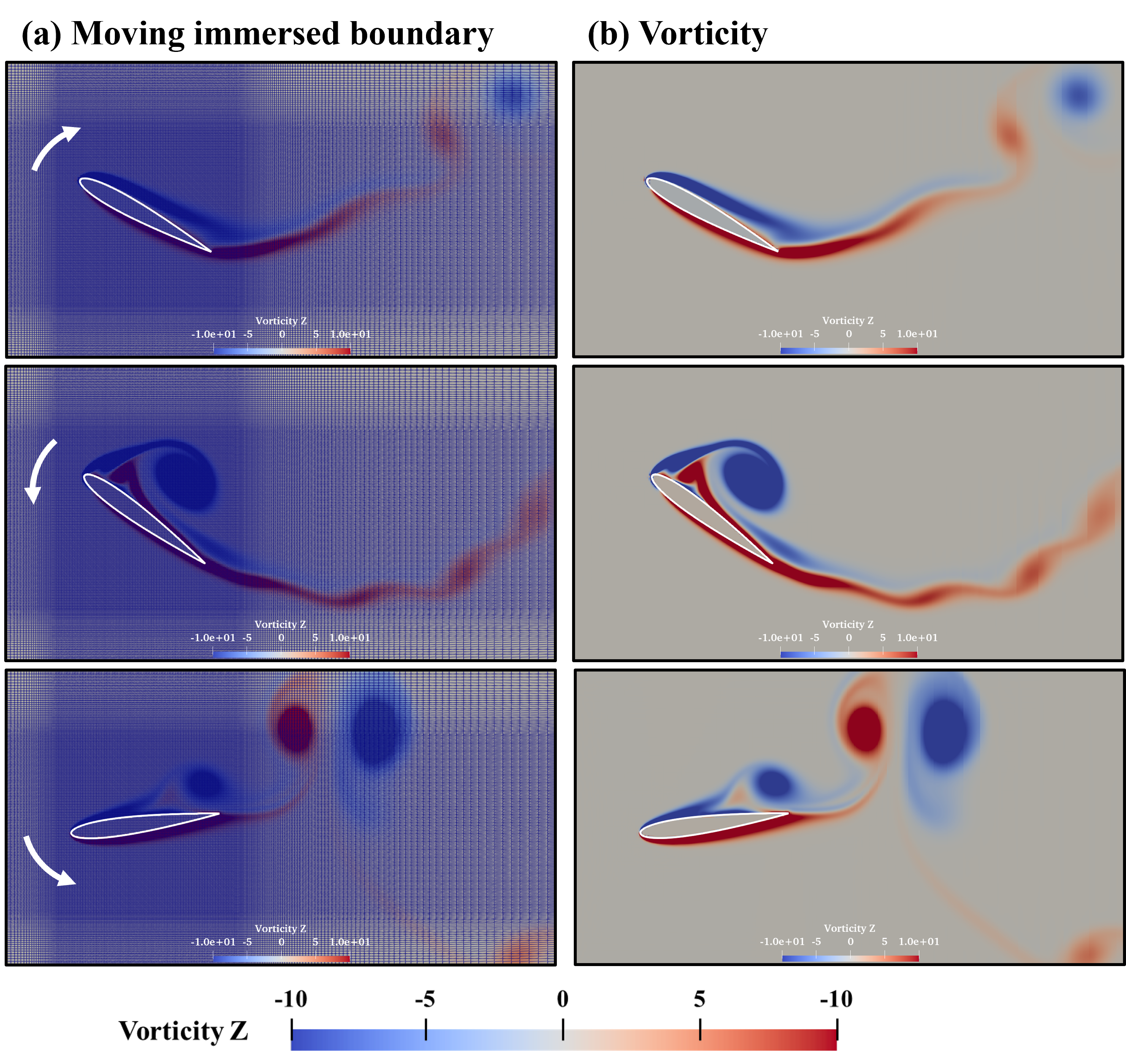} 
  \caption{Snapshots of the flow around a pitching airfoil with $p=3$: (a) IBM mesh and moving immersed boundary; (b) vorticity field showing unsteady vortex evolution.}
  \label{fig:Case1P-movingIB}
\end{figure}

The computational domain is $[-5c,\, 15c] \times [-10c,\, 10c]$. Seven Cartesian meshes with minimum element sizes $h = \Delta x_{\min} = \Delta y_{\min} \in \{0.015,\, 0.02,\, 0.03,\, 0.04,\, 0.05,\, 0.06,\, 0.08\}$ are used for the convergence study. A uniform mesh with spacing $h$ is employed in the near-body region 
$[-0.6c,\, 0.6c] \times [-0.6c,\, 0.6c]$. Outside this region, mesh stretching with a constant expansion ratio of $1.1$ is applied toward the far field to provide a smooth transition from the fine grid to the outer domain. For clarity in the convergence plots, the background mesh is considered as Cartesian (see Fig. \ref{fig:Case1P-movingIB}); nevertheless, the proposed DG-IBM implementation in HORSES3D is mesh-agnostic and readily applicable to unstructured hexahedral volume discretizations.

\begin{figure}[htbp]
  \centering
  \includegraphics[width=0.85\textwidth]{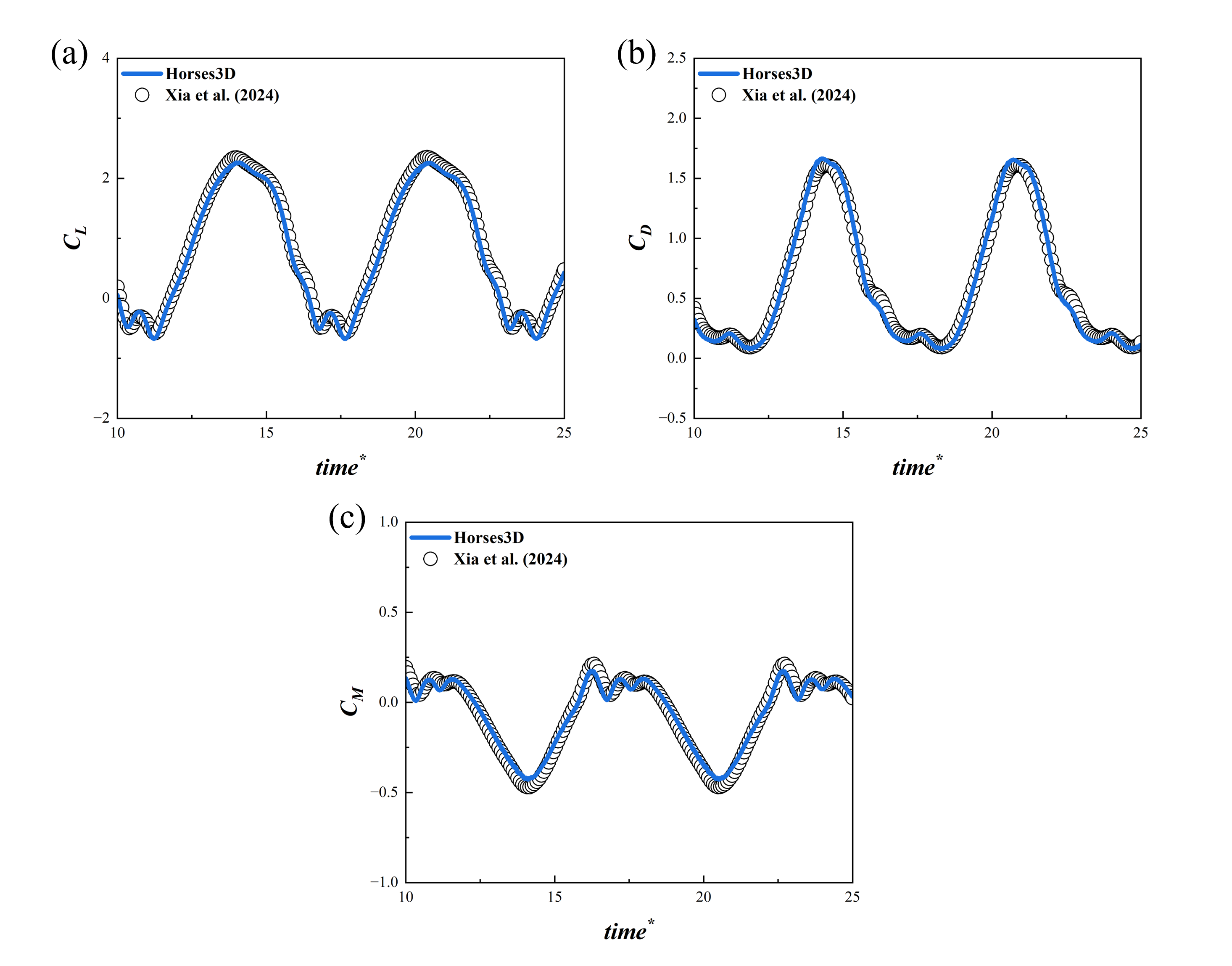} 
  \caption{Time histories of (a) $C_L$, (b) $C_D$, and (c) $C_M$ for the pitching airfoil compared with the reference data of Xia \textit{et al.} \cite{xia2024stall}.}
  \label{fig:Case1P-Validation}
\end{figure}

For 2D simulations, HORSES3D treats the domain as a 3D problem but uses only one element in the spanwise direction ($z$); this element can be assigned a low polynomial order since no physical variation is expected there. Consequently, we keep $p_z=1$ for all simulations; $p=3$ is used for fixed-$p$ computations, which therefore corresponds to $p_x=p_y=3,\;p_z=1$. The $p$-adaptation procedure adapts only $p_x$ and $p_y$ while maintaining $p_z=1$.

Fig. \ref{fig:Case1P-Validation} compares time histories of lift, drag and moment coefficients ($C_L$, $C_D$, $C_M$) against the reference data of Xia \textit{et al.} \cite{xia2024stall}, and good agreement is observed across the pitching cycles. Fig. \ref{fig:Case1P-movingIB} shows instantaneous flow fields, where panel (a) displays the moving immersed boundary and panel (b) gives the vorticity field for the $p=3$ run, illustrating complex vortex shedding induced by the large-amplitude motion.

\begin{figure}[htbp] % htbp
  \centering
  \includegraphics[width=0.85\textwidth]{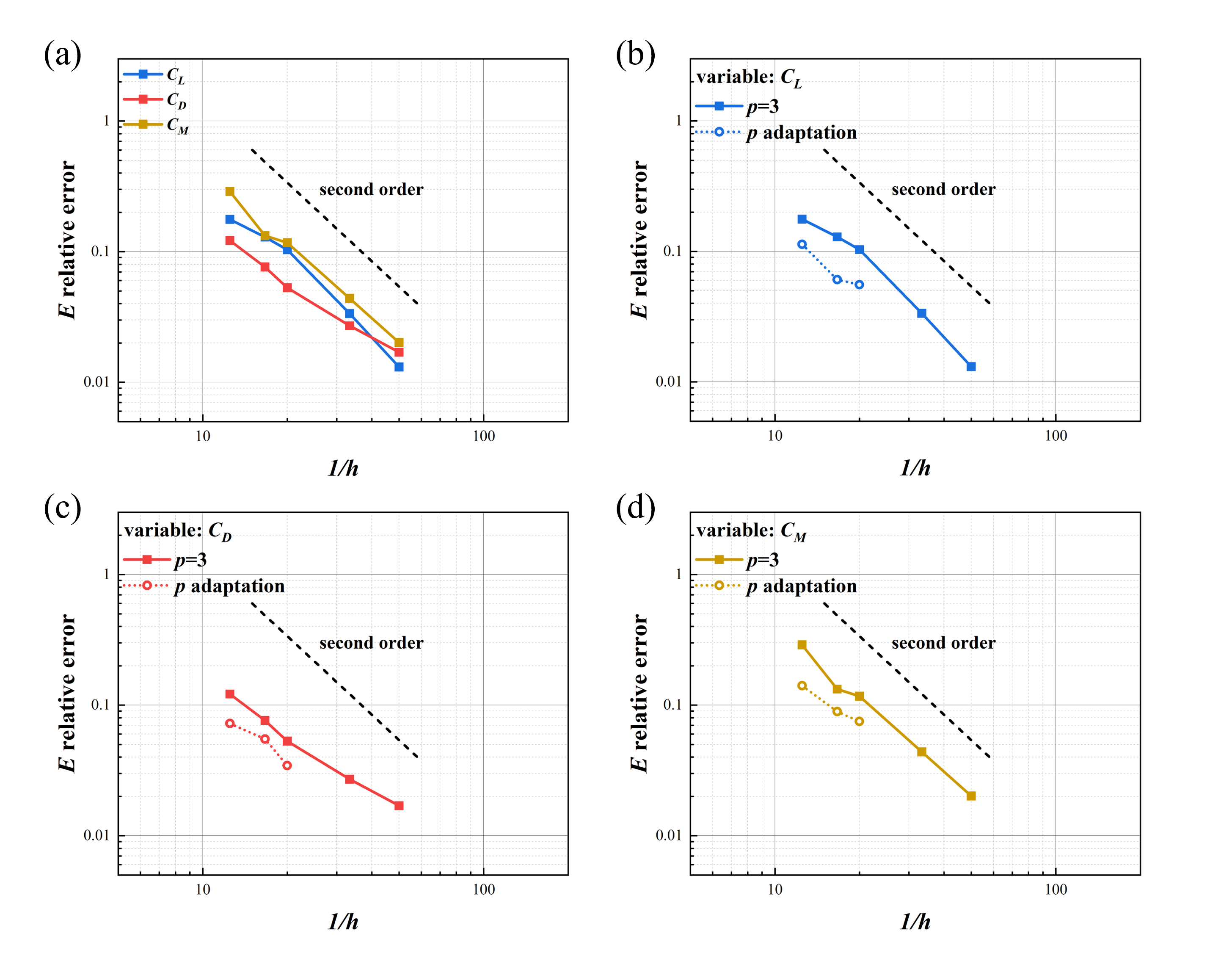} 
  \caption{Convergence and accuracy comparison for the pitching airfoil cases: (a) Relative error $E$ of $C_L$, $C_D$, and $C_M$ versus $1/h$ with uniform $p=3$; (b-d) force coefficient errors for the $p$-adaptation case compared with the uniform $p=3$ results.}
  \label{fig:Case1P-pConvergence}
\end{figure}

\begin{figure}[htbp] % htbp
  \centering
  \includegraphics[width=1\textwidth]{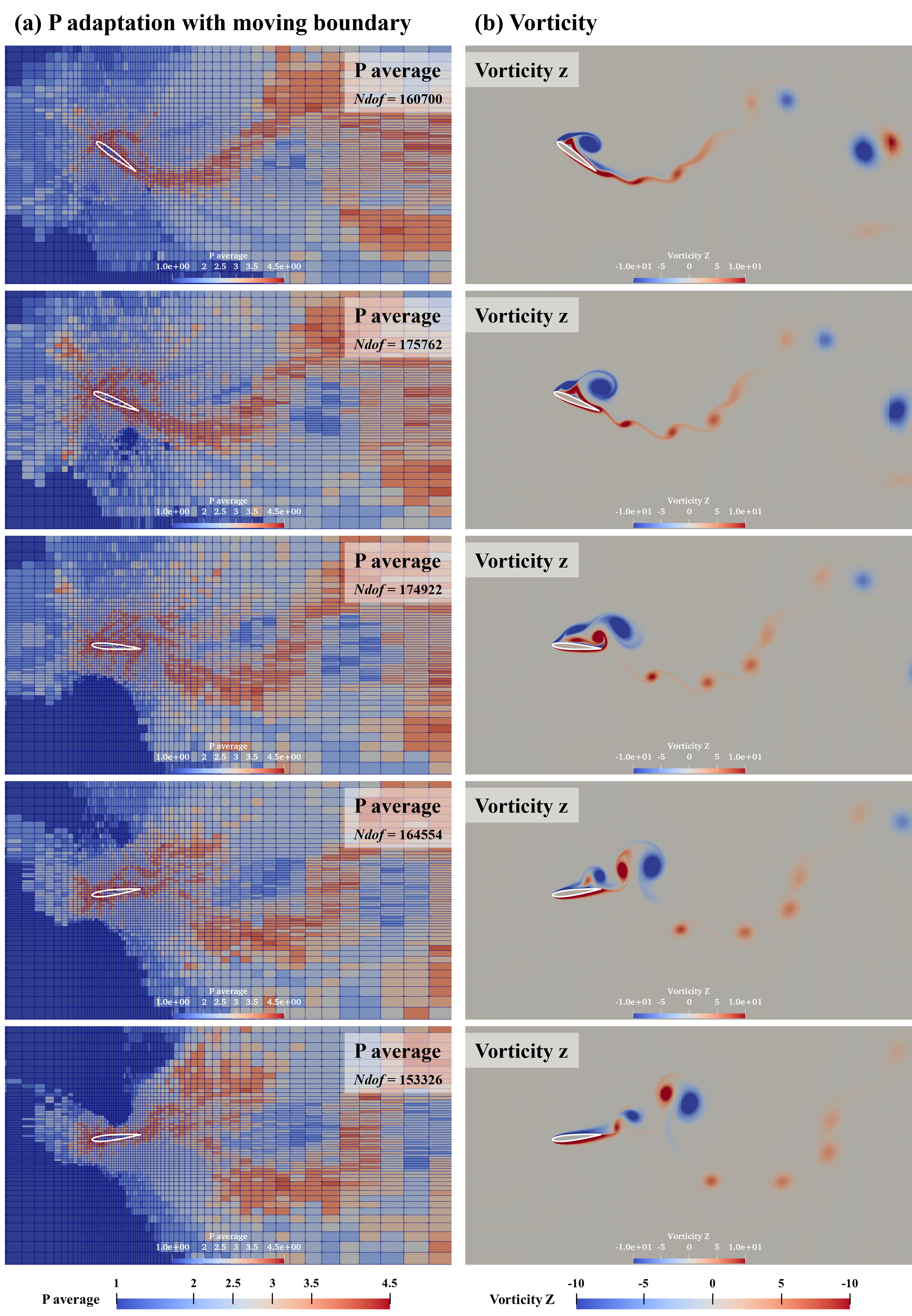} 
  \caption{Snapshots of the flow around a pitching airfoil with $p$-adaptation: (a) distribution of the average polynomial order in each element; (b) vorticity contours.}
  \label{fig:Case1P-pAdaptation}
\end{figure}

A convergence study is conducted by measuring the relative errors $E$ of $C_L$, $C_D$ and $C_M$ over several (typically 2-3 cycles) selected pitching cycles $\tau$. Given a coefficient time history $C(t)$ and a reference signal 
$C_{\mathrm{ref}}(t)$, the relative error is defined as
\begin{equation}
E(C) =
\frac{
\left( \int_{\tau} \left[ C(t)-C_{\mathrm{ref}}(t) \right]^2 dt \right)^{1/2}
}{
\left( \int_{\tau} C_{\mathrm{ref}}(t)^2 dt \right)^{1/2}
},
\label{eq:relative-error}
\end{equation}
This metric captures the discrepancy in the full temporal evolution of the coefficient and is appropriate for unsteady periodic fluid dynamics. The reference solution corresponds to the finest grid with $h=0.015$ and $p=3$. 
It reveals that the relative errors of $C_L$, $C_D$ and $C_M$ decay approximately with second-order accuracy as $h$ is reduced as shown in Fig. \ref{fig:Case1P-pConvergence}(a). Furthermore, according to Fig. \ref{fig:Case1P-pConvergence}(b-d), activating the RL-driven anisotropic $p$-adaptation results in smaller relative errors than the uniform $p=3$ cases, while reducing the total number of degrees of freedom $Ndof$ as shown in Fig. \ref{fig:Case1P-pAdaptation}(a). For the test case shown in Fig. \ref{fig:Case1P-pAdaptation}, the uniform $p=3$ corresponds to $Ndof=204\,672$, and the adaptive run obtains a better accuracy with a smaller $Ndof=160\,700$ approximately on average. 

To display the results, the spatial average of the local polynomial order is computed in each element as
\begin{equation}
P_{\mathrm{average}}=\frac{p_x+p_y+p_z}{3},
\end{equation}
and shown in Fig. \ref{fig:Case1P-pAdaptation}(a). Compared to the vorticity snapshots shown in Fig. \ref{fig:Case1P-pAdaptation}(b), it demonstrates that the RL-driven agent increases $p$ in the vicinity of the moving immersed boundary and in regions of vortex development. In particular, the adaptive scheme raises $p$ where the mask transition and vortex interaction occur, and lowers $p$ when these features advect away or weaken, thus producing a time-varying, cost-aware resolution distribution.

In summary, this test demonstrates that (i) the symmetric Gaussian quadrature on triangles of immersed boundary, combined with IDW surface reconstruction, results in accurate fluid-force estimates for large-amplitude moving boundaries; (ii) the RL-driven $p$-adaptation effectively concentrates resolution where required by the unsteady flow and the moving boundary, reducing computational cost while increasing accuracy.

\subsection{Stall flutter of an airfoil}
\label{sec: test case - Stall flutter}

This test simulates a classic stall flutter problem to evaluate the predictive capability of the partitioned FSI coupling strategy with rigid-body motion. The settings of the test cases are identical to those used in Section \ref{sec:Case1P}. The difference with respect to the previous section is that before the airfoil motion was prescribed, while now the motion is calculated by the FSI framework.

The rigid-body structural model follows the formulations in Section \ref{sec:structural solver}(a). Relevant nondimensional parameters are: the dimensionless moment of inertia $I^*=I/(0.5\,\rho c^4)=4.5$; the damping ratio $\zeta^*=D/D_{\mathrm{cr}}=0.03$, where $D_{\mathrm{cr}}=2\sqrt{K I_{\mathrm{ea}}}$ is the critical damping for the harmonic oscillator; and a reduced velocity $U^*=U_\infty/f_sc=6$, where $f_s=(1/2\pi)\sqrt{K/I_{\mathrm{ea}}}$) is the natural frequency of the spring.

\begin{figure}[htbp] % htbp
  \centering
  \includegraphics[width=1\textwidth]{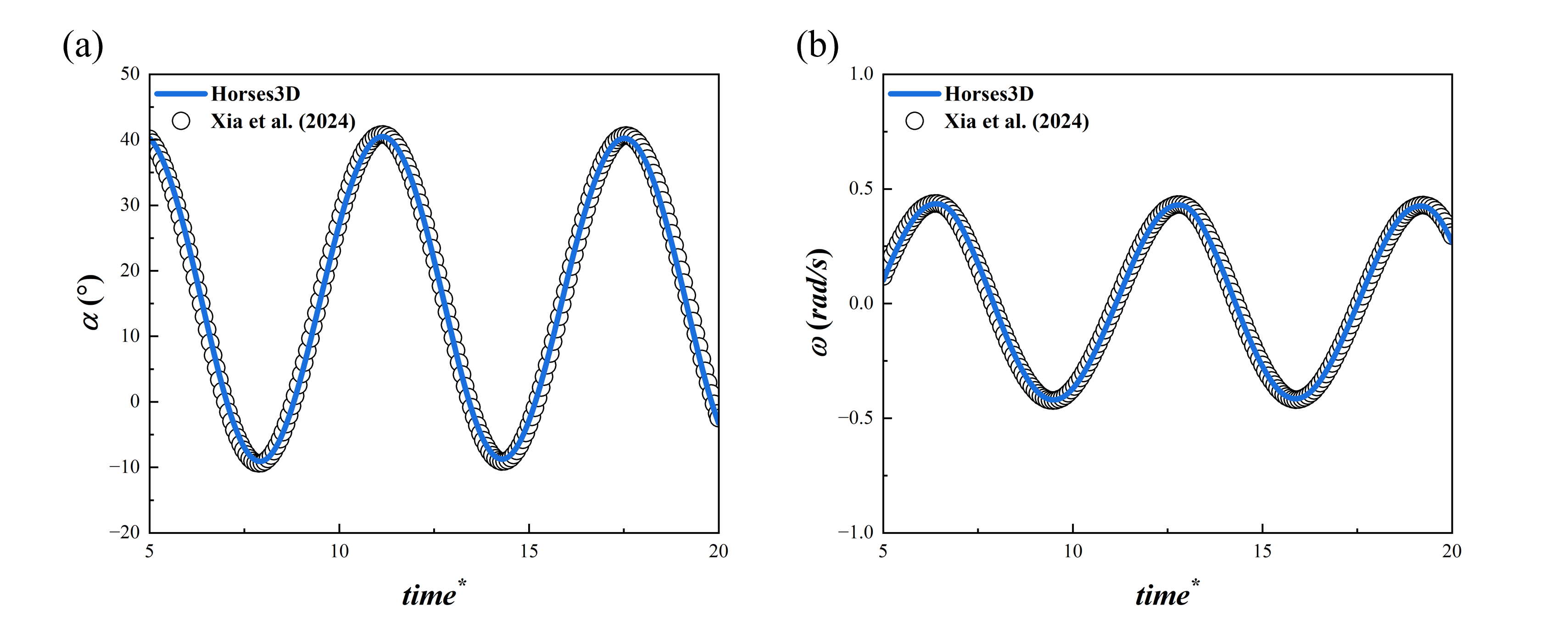} 
  \caption{Time histories of (a) angle of attack $\alpha$, (b) angular velocity $\omega$ for the airfoil stall flutter compared with the reference data of Xia \textit{et al.} \cite{xia2024stall}.}
  \label{fig:Case1F-Validation}
\end{figure}

\begin{figure}[htbp] % htbp
  \centering
  \includegraphics[width=0.8\textwidth]{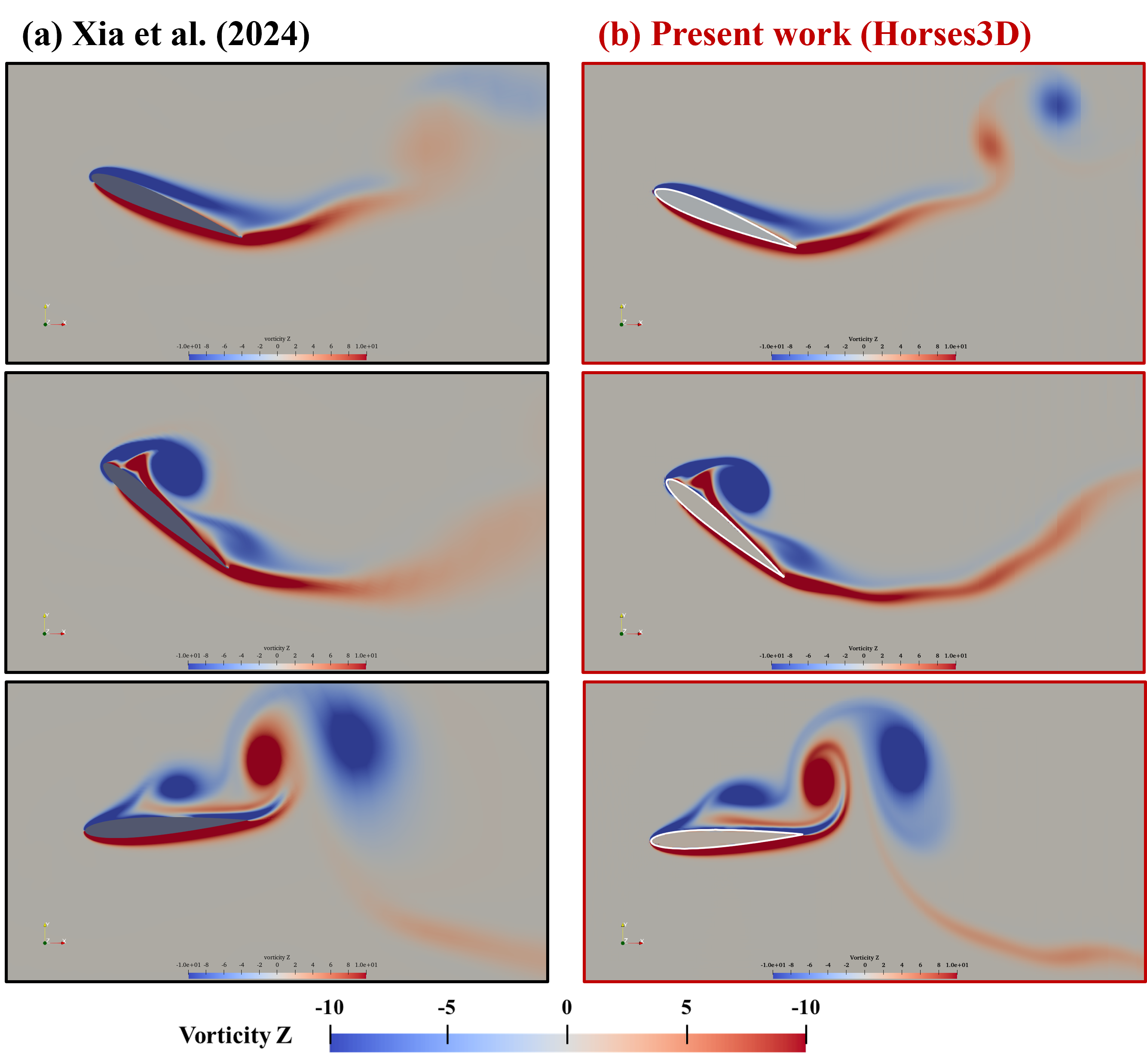} 
  \caption{Snapshots of the vorticity field for the airfoil stall flutter compared with Xia \textit{et al.} \cite{xia2024stall}.}
  \label{fig:Case1F-vorticity}
\end{figure}

\begin{figure}[htbp] % htbp
  \centering
  \includegraphics[width=0.7\textwidth]{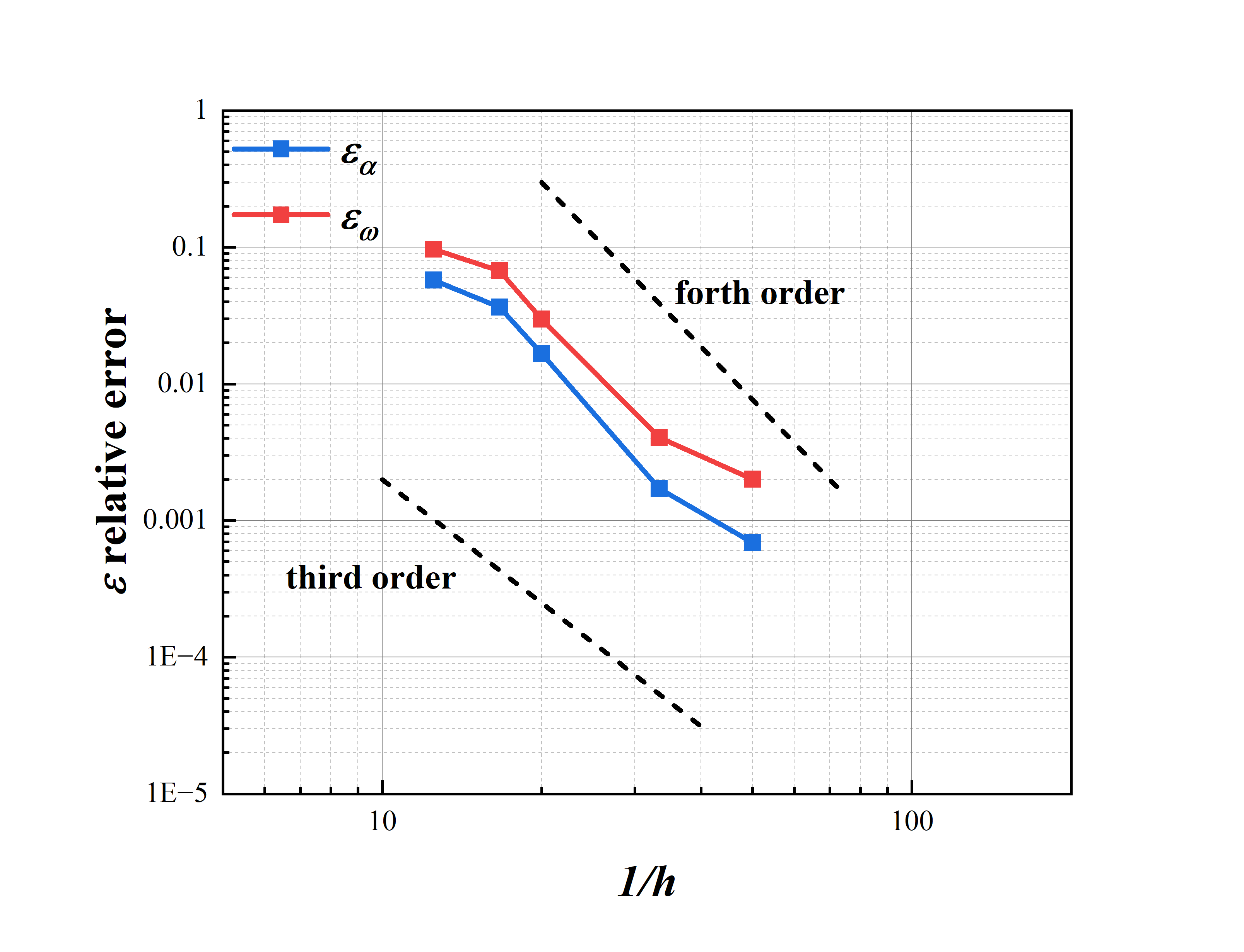} 
  \caption{Convergence and accuracy comparison for the airfoil stall flutter cases: relative errors $\varepsilon_\alpha$ and $\varepsilon_\omega$ versus $1/h$ for uniform $p=3$. Here $\varepsilon_\alpha=(\alpha_{max}^{ref}-\alpha_{max})/\alpha_{max}^{ref}$ and $\varepsilon_\omega=(\omega_{max}^{ref}-\omega_{max})/\omega_{max}^{ref}$.}
  \label{fig:Case1F-pConvergence-p3}
\end{figure}

\begin{figure}[htbp] % htbp
  \centering
  \includegraphics[width=1\textwidth]{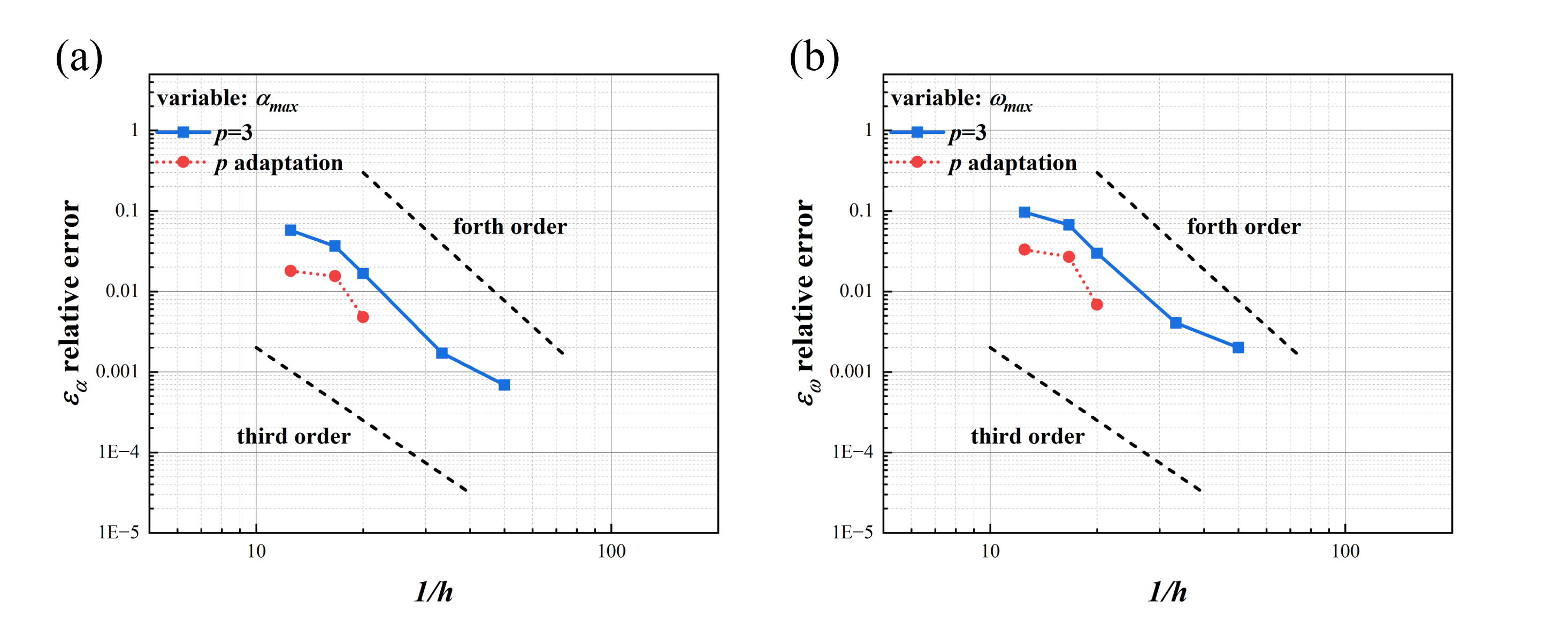} 
  \caption{Relative errors of (a) $\varepsilon_\alpha$ and (b) $\varepsilon_\omega$ for $p$-adaptation cases compared with the uniform $p=3$ results.}
  \label{fig:Case1F-pAdaptError}
\end{figure}

\begin{figure}[ht] % htbp
  \centering
  \includegraphics[width=1\textwidth]{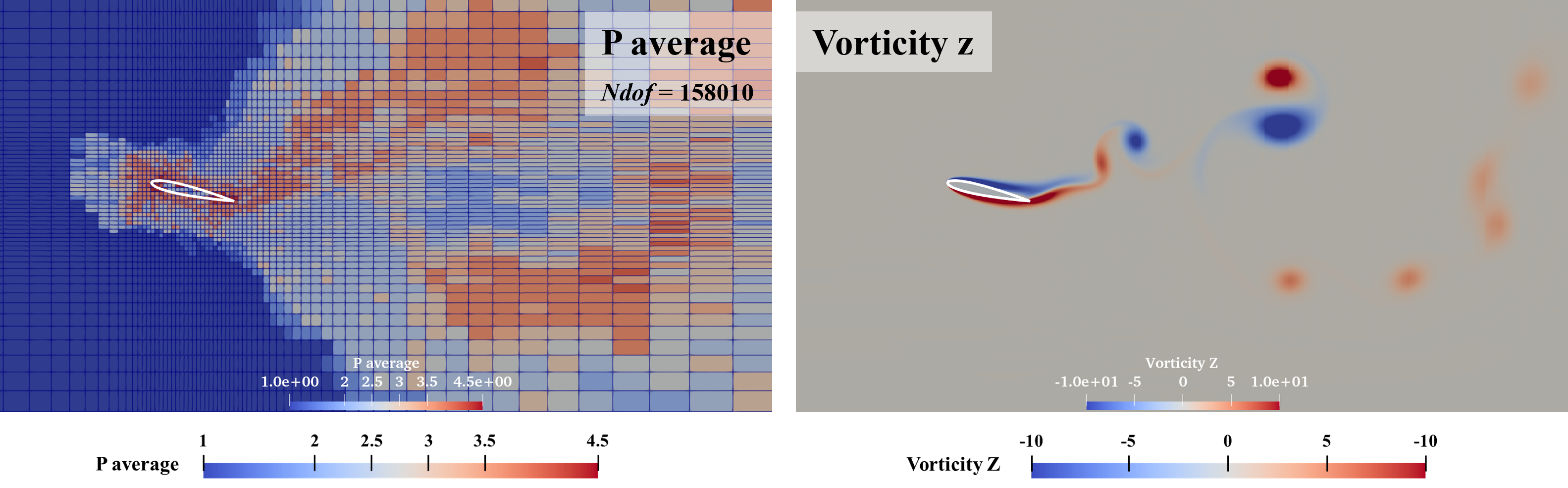} 
  \caption{Snapshots of the airfoil stall flutter case with $p$-adaptation: (a) distribution of the average polynomial order in each element; (b) vorticity contours.}
  \label{fig:Case1F-pAdaptation}
\end{figure}

Fig. \ref{fig:Case1F-Validation} displays the time histories of the angle of attack $\alpha(t)$ and the angular velocity $\omega(t)=\dot{\alpha}(t)$ compared with the reference data of Xia \textit{et al.}~\cite{xia2024stall}. Good agreement is observed in both transient and limit-cycle behavior. Fig. \ref{fig:Case1F-vorticity} further presents representative vorticity snapshots. Qualitative features of the separated flow and vortex dynamics match the reference results well.

To quantify the convergence of structural responses \cite{kim2018weak}, we define the relative errors for the peak quantities as 
\begin{equation}
\varepsilon_{\alpha}=\frac{\alpha_{\max}^{\mathrm{ref}}-\alpha_{\max}}{\alpha_{\max}^{\mathrm{ref}}},\qquad
\varepsilon_{\omega}=\frac{\omega_{\max}^{\mathrm{ref}}-\omega_{\max}}{\omega_{\max}^{\mathrm{ref}}},
\end{equation}
where $\alpha_{\max}$ and $\omega_{\max}$ denote the maximum pitch angle and maximum angular velocity obtained on the current mesh, and the superscript $ref$ refers to the finest-mesh solution ($h=0.015$). Fig. \ref{fig:Case1F-pConvergence-p3} plots $\varepsilon_{\alpha}$ and $\varepsilon_{\omega}$ versus $1/h$ for uniform $p=3$. The decay of these errors with mesh refinement indicates that the coupled solver attains observed convergence rates approaching third to fourth order for the structural peak quantities.

The effect of the RL-driven anisotropic $p$-adaptation is examined next. Fig. \ref{fig:Case1F-pAdaptError} compares the relative errors $\varepsilon_{\alpha}$ and $\varepsilon_{\omega}$ for the $p$-adaptation against the uniform polynomial $p=3$. It shows that the adaptive strategy reduces both errors across the tested meshes. Fig. \ref{fig:Case1F-pAdaptation} shows a representative snapshot for the $p$-adaptive run: panel (a) displays the element-wise average polynomial order $P_{\mathrm{average}}$, and panel (b) shows the corresponding vorticity contours. For this example the adaptive simulation achieves the reported accuracy with $Ndof=158\,010$, substantially fewer than the uniform $p=3$ case ($Ndof=204\,672$), demonstrating that the $p$-adaptation achieves both higher accuracy and reduced computational cost for the coupled rigid FSI problem.

In summary, the stall flutter case confirms that the partitioned DG-IBM coupling accurately predicts rigid-body aeroelastic responses; moreover, the RL-based $p$-adaptation improves efficiency and accuracy for FSI responses.

\subsection{Flow-induced vibration of an elastic beam behind a cylinder}
\label{sec: test case - elastic beam}

This last case examines the predictive capability of the high-order DG-IBM framework coupling with the elastic structural solver, and the performance of the $p$-adaptation strategy involving flexible deformation. 

\begin{figure}[htbp] % htbp
  \centering
  \includegraphics[width=0.7\textwidth]{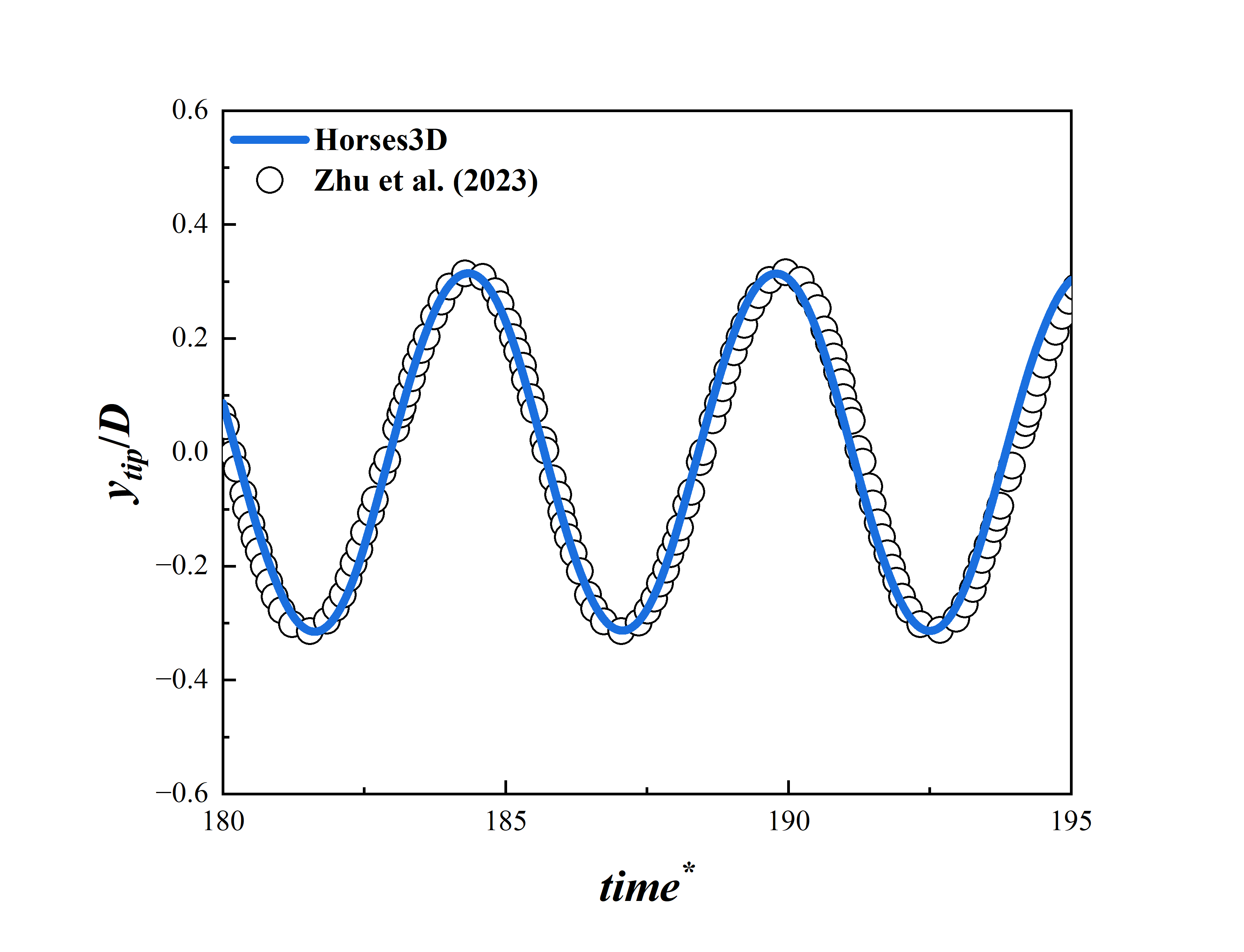} 
  \caption{Time histories of $y_\mathrm{tip}$ for the elastic beam behind a cylinder compared with the reference data of Zhu \textit{et al.} \cite{zhu2023flow}.}
  \label{fig:Case2-Validation}
\end{figure}

\begin{figure}[htbp] % htbp
  \centering
  \includegraphics[width=0.7\textwidth]{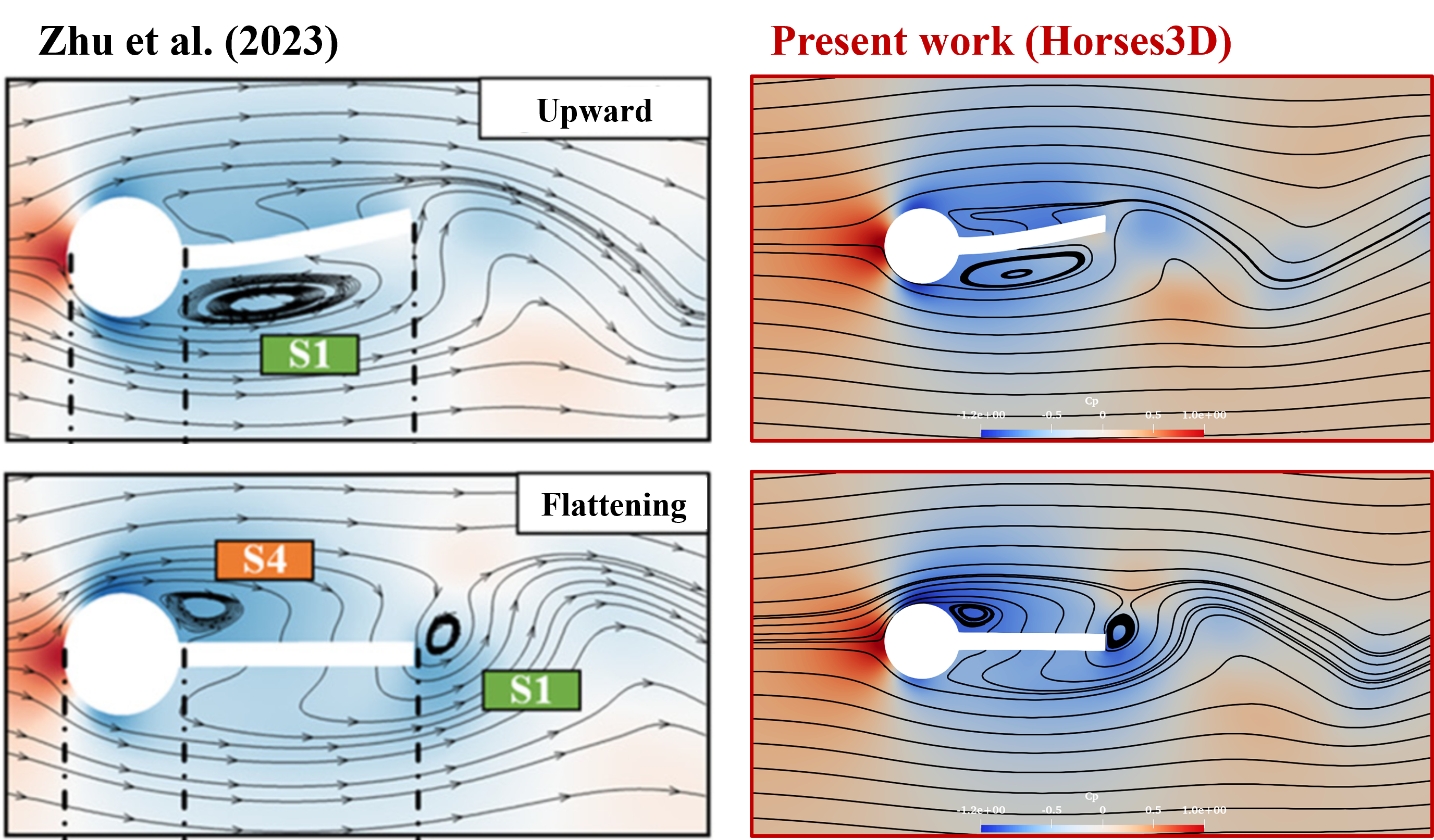} 
  \caption{Snapshots of the pressure coefficient $C_p$ and streamlines for the elastic beam behind a cylinder compared with the reference data of Zhu \textit{et al.} \cite{zhu2023flow}.}
  \label{fig:Case2-CpField}
\end{figure}

The benchmark corresponds to the classical configuration of a rigid circular cylinder followed by an elastic beam, as studied by Zhu \textit{et al.}~\cite{zhu2023flow}. The setup involves a uniform free-stream flow past a stationary circular cylinder of diameter $D$ with an attached elastic beam of length $L/D = 2.0$ and thickness $H/D = 0.2$. The density ratio is $\beta = \rho_s / \rho_f = 1.0$, the elastic modulus $A_E = E_s / (\rho_f U^2) = 2187.5$. The first three structural modes are used, with the natural frequencies $fD/U=0.41, 2.43, 6.39$. The Reynolds number based on $D$ is $Re = 160$, and the Mach number is $Ma = 0.25$. The computational domain extends over $[-10D, 25D] \times [-10D, 10D]$. The CFL number is set to 0.3, and the polynomial order is fixed at $p=3$.

% The spatial discretizations consist of six Cartesian meshes with minimum element sizes $h = \Delta x_{min} = \Delta y_{min} \in \{0.05,\,0.054,\,0.056,\,0.06,\,0.08,\,0.15\}$. The finest mesh ($h=0.05$) is employed as the reference. 
The displacement at the tip of the elastic beam, denoted as $y_\mathrm{tip}$, is monitored during the computation and used as the primary response indicator. The structural dynamics are solved using the elastic solver described in Section \ref{sec:structural solver}(b), and the fluid field is computed by the DG-IBM framework introduced in Section \ref{sec:Numerical Methodology}.

Fig. \ref{fig:Case2-Validation} presents the time history of $y_\mathrm{tip}(t)$ compared with the reference results of Zhu \textit{et al.}~\cite{zhu2023flow}. The tip displacement shows great agreement with the literature, confirming the fidelity of the coupled elastic solver. Fig. \ref{fig:Case2-CpField} illustrates representative snapshots of the pressure coefficient $C_p$ and streamlines. Both the flow pattern and the instantaneous deformation of the flexible beam closely resemble the reference data.

% To assess the numerical accuracy, we define relative errors \cite{kim2018weak} for the maximum displacement and velocity at the beam tip as
% \begin{equation}
% \varepsilon_{y_\mathrm{tip}} = \frac{y_{\mathrm{tip},\max}^{\mathrm{ref}} - y_{\mathrm{tip},\max}}{y_{\mathrm{tip},\max}^{\mathrm{ref}}}, \qquad
% \varepsilon_{v_\mathrm{tip}} = \frac{v_{\mathrm{tip},\max}^{\mathrm{ref}} - v_{\mathrm{tip},\max}}{v_{\mathrm{tip},\max}^{\mathrm{ref}}}.
% \end{equation}
% Fig. \ref{fig:Case2-pConvergence-p3} shows the variation of these relative errors with $1/h$ for the uniform $p=3$ simulations (solid lines). The results demonstrate overall convergence rates close to third to fourth order, indicating the robustness of the DG-IBM coupling in capturing FSI with elastic deformation. The corresponding $p$-adaptive results are also plotted as open circles, revealing a substantial reduction in both displacement and velocity errors compared to the uniform-$p$ baseline on the same meshes.

% \begin{figure}[htbp] % htbp
%   \centering
%   \includegraphics[width=0.7\textwidth]{images/Case2-pConvergence-p3.png} 
%   \caption{Convergence and accuracy comparison for the elastic beam behind a cylinder case: solid lines represent relative errors $\varepsilon_{y_\mathrm{tip}}$ and $\varepsilon_{v_\mathrm{tip}}$ versus $1/h$ for uniform $p=3$; open circles indicate relative errors for the $p$-adaptation cases.}
%   \label{fig:Case2-pConvergence-p3}
% \end{figure}

\begin{figure}[htbp] % htbp
  \centering
  \includegraphics[width=0.65\textwidth]{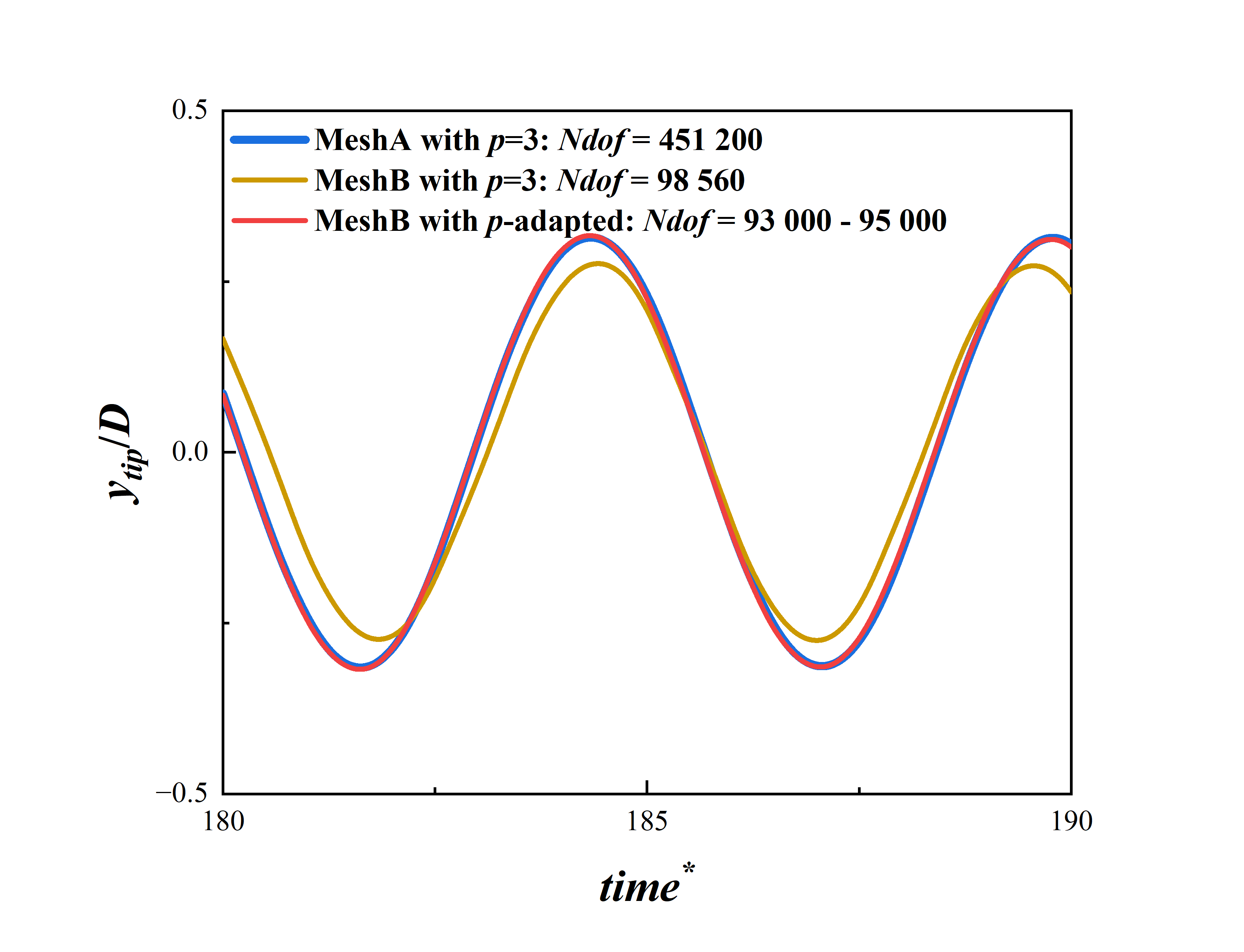} 
  \caption{Comparison for the elastic beam behind a cylinder case for uniform $p=3$ (and two h-meshes) and $p$-adaptation cases.}
  \label{fig:Case2-pAdaptation-ytip}
\end{figure}

% Panel (a) shows the elementwise distribution of the average polynomial order $P_\mathrm{average}$, with the adaptive simulation achieving $Ndof$ slightly lower than the uniform $p=3$ case ($Ndof=98\,560$). Panels (b) and (c) display the instantaneous contours of the velocity components in $x$-direction and $y$-direction, respectively. It is observed that the RL-driven agent, in addition to capturing the moving immersed boundary, also distributed a large number of fluid elements with high $p$ at the wake. As a result, the adaptive approach leads to a small reduction in total degrees of freedom,
% % but a significant improvement in overall prediction accuracy, 
% demonstrating its effectiveness for FSI problems with flexible structures.

The effectiveness of the anisotropic $p$-adaptation strategy is further examined in this case, as shown in Fig.~\ref{fig:Case2-pAdaptation-ytip}. The fine reference mesh (Mesh~A, $\Delta x_{min}=\Delta y_{min}=0.05$, $p=3$, $Ndof=451\,200$) reproduces the benchmark solution reported in the literature, as shown in Fig.\ref{fig:Case2-Validation}. In the coarse mesh (Mesh~B, $\Delta x_{min}=\Delta y_{min}=0.15$), a uniform polynomial order of $p=3$ ($Ndof=98\,560$) results in a noticeable deviation in the predicted tip displacement. When $p$-adaptation is activated, the agent assigns higher orders near the moving boundary and in dynamically evolving flow regions, resulting in a solution that closely matches Mesh~A while only requiring $Ndof = 93\,000\text{-}95\,000$, which is approximately one-fifth of the fine-mesh cost. This highlights the substantial efficiency improvement achieved without compromising accuracy.

\begin{figure}[htbp] % htbp
  \centering
  \includegraphics[width=1\textwidth]{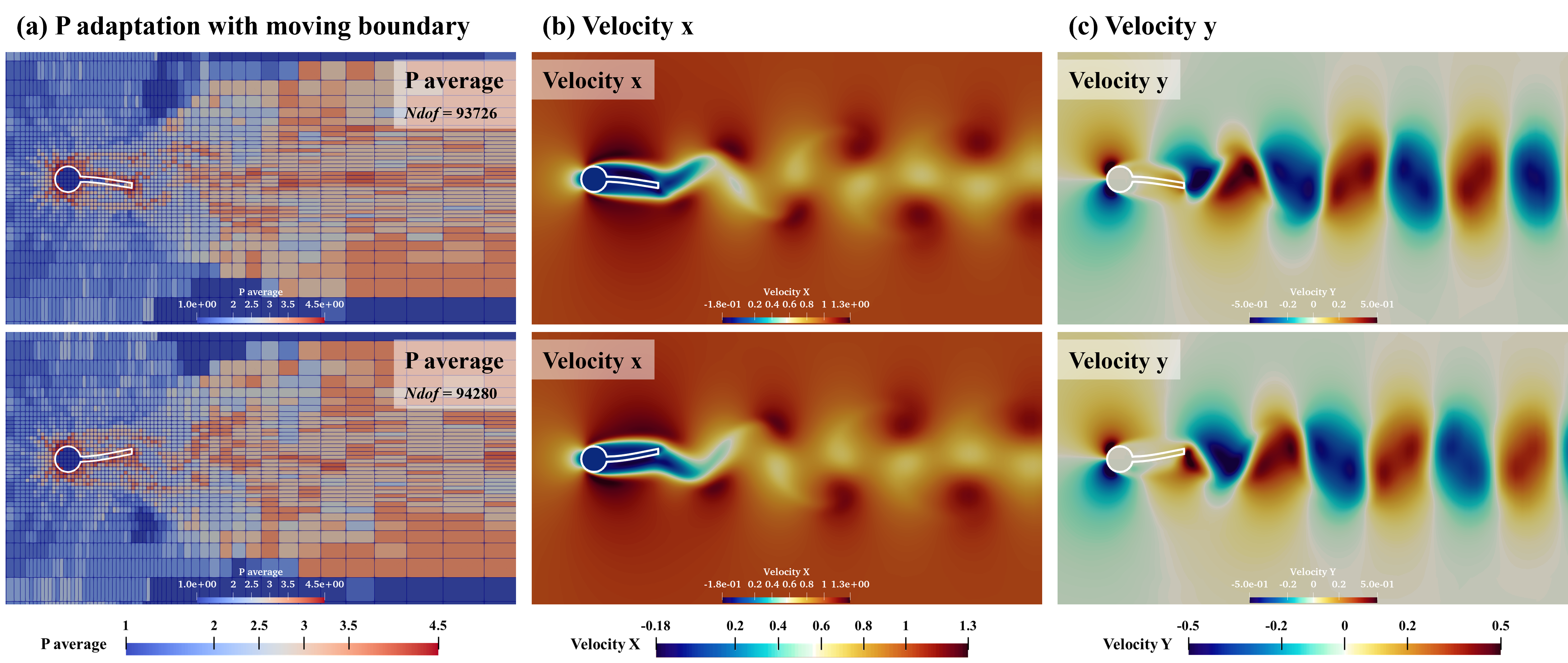} 
  \caption{Snapshots of the elastic beam behind a cylinder case with $p$-adaptation: (a) distribution of the average polynomial order in each element; (b) contours of velocity in $x$-direction $u$; (c) contours of velocity in $y$-direction $v$.}
  \label{fig:Case2-pAdaptation}
\end{figure}

Figure~\ref{fig:Case2-pAdaptation} shows the adapted field: panel~(a) presents the elementwise average polynomial order $P_{\mathrm{average}}$, while panels~(b) and~(c) display instantaneous contours of the velocity components in the $x$- and $y$-directions. The RL-agent consistently increases the polynomial resolution not only around the moving immersed boundary but also within the unsteady wake. Compared with the uniform-$p$ Mesh~B, the adaptive solution achieves markedly improved predictive fidelity despite a comparable number of degrees of freedom, demonstrating its effectiveness for FSI problems involving flexible structures.

\section{Conclusions}
\label{sec:Conclusions}

% A high-order discontinuous Galerkin-immersed boundary framework for fluid-structure interaction is developed. The method couples a nodal DG solver with a volume-penalization immersed boundary formulation for moving interfaces, employs symmetric high-order Gaussian quadrature for accurate interfacial force evaluation, and integrates rigid-body and elastic structural solvers within a partitioned coupling strategy. An anisotropic p-adaptation procedure guided by a reinforcement-learning agent dynamically refines polynomial resolution near moving boundaries and vortical regions, achieving a superior balance between accuracy and computational cost. Validation on several benchmark problems (flow past a pitching airfoil, stall flutter, and flow-induced vibration of an elastic beam) demonstrates the high-order accuracy, high efficiency and robustness of the proposed method. Future work will extend it to three-dimensional turbulent flows and complex moving geometries.

A high-order discontinuous Galerkin-immersed boundary framework has been developed for simulating fluid-structure interaction with moving immersed boundaries. The method couples a nodal DG solver with a volume-penalization immersed boundary formulation, and employs symmetric high-degree Gaussian quadrature for accurate evaluation of interfacial loads. Rigid-body and elastic structures are integrated within a partitioned strategy using analytical and modal solvers, ensuring robustness for fully coupled motions and adaptability to the volume-penalization immersed boundary method.
To improve efficiency while maintaining high-order accuracy, an anisotropic reinforcement-learning-based $p$-adaptation procedure is introduced to dynamically adjust local polynomial orders in response to boundary motion and flow features. 

Validations on a pitching airfoil, stall flutter, and flow-induced vibration of an elastic plate demonstrate high-order accuracy, robust fluid-structure coupling, and substantial efficiency gains resulting from $p$-adaptation. In general, the results confirm that the proposed DG-IBM-$p$-adaptation framework provides a practical, high-fidelity approach for complex moving-boundary FSI problems. Future work will extend it to three-dimensional turbulent flows and complex moving geometries.

\section{Acknowledgements}
This research has received funding from the European Union (ERC, Off-coustics, project number 101086075). Views and opinions expressed are, however, those of the authors only and do not necessarily reflect those of the European Union or the European Research Council. Neither the European Union nor the granting authority can be held responsible for them.

YX thanks the support of the China Scholarship Council (Project ID: 202306020136).

EF acknowledges the funding received by the Grant DeepCFD (Project No. PID2022-137899OB-I00) funded by MICIU/AEI/10.13039/501100011033 and by ERDF, EU.

DH and EF acknowledge the funding received by the Comunidad de Madrid according to Orden 5067/2023, of December 27th, issued by the Consejero de Educación, Ciencia y Universidades, which announces grants for the hiring of predoctoral research personnel in training for the year 2023.

We thankfully acknowledge the computer resources at MareNostrum and the technical support provided by the Barcelona Supercomputing Center (RES-IM-2025-3-0002).
All authors gratefully acknowledge the Universidad Politécnica de Madrid (www.upm.es) for providing computing resources on the Magerit Supercomputer.

\begin{appendices}
\section{Sensitivity of the immersed-boundary force to \texorpdfstring{$N_{\rm IP}$}{NIP}}
\label{sec:Appendix-A}

\begin{figure}[ht] % htbp
  \centering
  \includegraphics[width=1\textwidth]{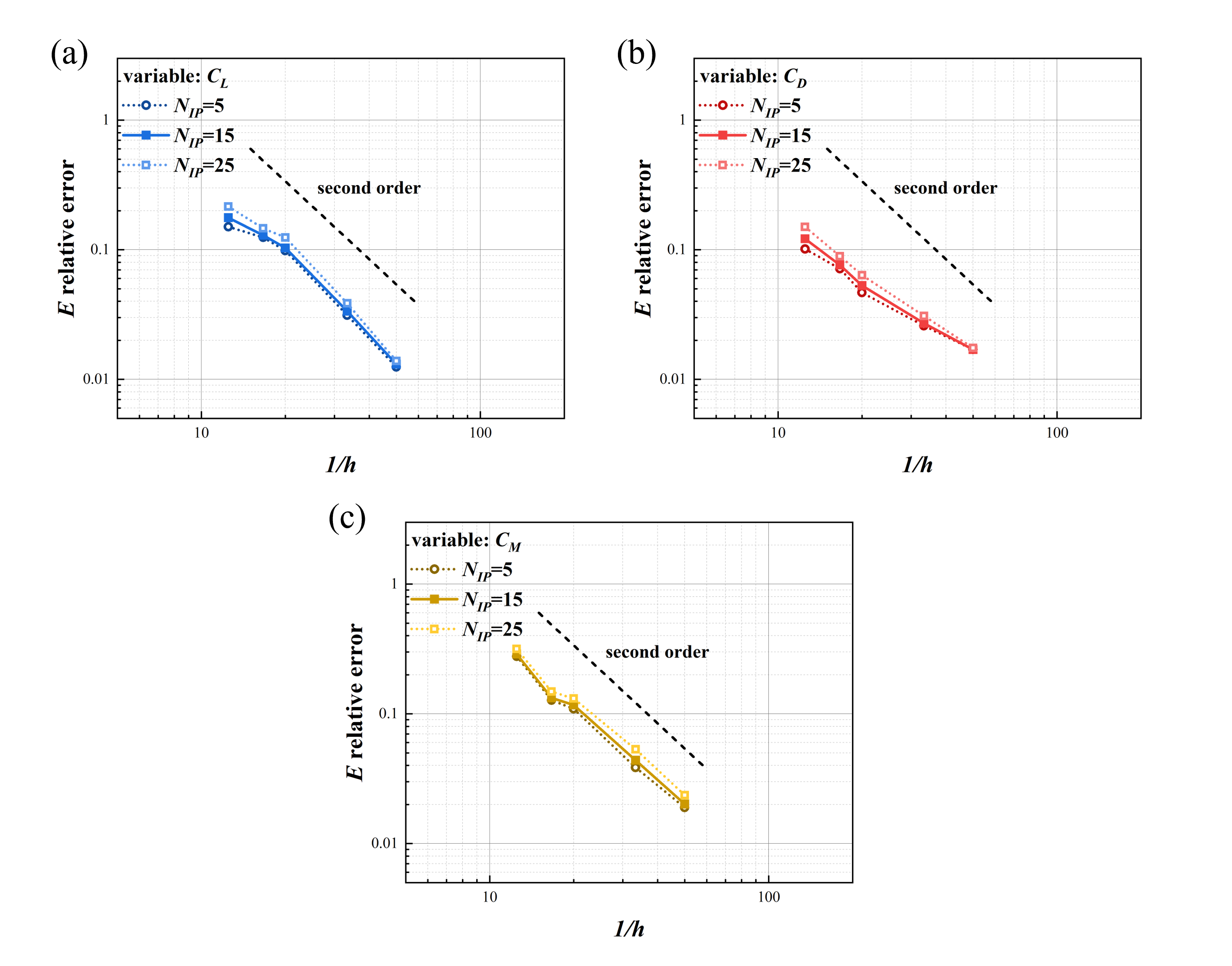} 
  \caption{Convergence and accuracy comparison for the pitching airfoil cases: behavior of the error as ${N_{\rm IP}}$ changes.}
  \label{fig:AppendixA-Case1-checkN}
\end{figure}

This appendix examines the influence of the number of interpolation points, $N_{\rm IP}$, used in the IDW reconstruction of fluid quantities on the immersed boundary. The analysis is performed using the pitching airfoil case introduced in Section~\ref{sec:Case1P}. Figure~\ref{fig:AppendixA-Case1-checkN} presents the convergence behavior of the immersed boundary force errors for $N_{\rm IP}=5$, $15$, and $25$.

The results indicate that for coarse meshes, smaller values of $N_{\rm IP}$ lead to slightly slower convergence due to the increased sensitivity of the IDW procedure to local solution irregularities. However, as the mesh is refined, the convergence trends for all three values of $N_{\rm IP}$ become nearly indistinguishable. This study shows that the choice of $N_{\rm IP}$ does not alter the overall convergence order of the solver.

In addition, increasing $N_{\rm IP}$ introduces additional smoothing in the reconstructed boundary quantities, which results in a modest degradation in accuracy. Balancing the trade-off between convergence behavior and accuracy, the present work adopts $N_{\rm IP}=15$, which provides stable convergence while maintaining high predictive accuracy in all mesh resolutions.

\end{appendices}

\bibliographystyle{unsrt}
\bibliography{references.bib}

\end{document}